\documentclass[aps,prl,showpacs,twocolumn,groupedaddress]{revtex4}  % for submission
\usepackage{mya4,amsfonts,amssymb,epsfig}

\RequirePackage[usenames]{pstcol}

\def\MET{{\mbox{$E\kern-0.57em\raise0.19ex\hbox{/}_{T}$}}}
\def\met{{\mbox{$E\kern-0.57em\raise0.19ex\hbox{/}_{T}$}}}

\def\ifb{fb$^{-1}$}

\def\WH{$WH\rightarrow \ell\nu b\bar{b}$}
\def\whe{$WH\rightarrow e\nu b\bar{b}$}
\def\whm{$WH\rightarrow \mu\nu b\bar{b}$}
\def\lmet{$WH\rightarrow \ell\kern-0.45em\raise0.19ex\hbox{/} \nu b\bar{b}$}

\def\zhv{$ZH\rightarrow \nu\bar{\nu} b\bar{b}$}

\def\zhee{$ZH\rightarrow eeb\bar{b}$}

\def\zhmm{$ZH\rightarrow \mu \mu b\bar{b}$}
\def\zhl{$ZH\rightarrow \ell^+ \ell^- b\bar{b}$}

\def\pzhh{$p\bar{p}\rightarrow ZH \rightarrow  \nu\bar{\nu} b\bar{b} / \ell \ell b\bar{b}$}

\def\pwwww{$p\bar{p}\rightarrow WH \rightarrow WW W$}

\def\wwww{$WH \rightarrow WWW$}

\def\phwww{$p\bar{p}\rightarrow H \rightarrow WW$}

\def\hwww{$H\rightarrow WW$}
\def\hbb{$H\rightarrow b\bar{b}$}

\def\gevc{~GeV}

\begin{document}   

\newcommand{\MEx}{$\not\hspace*{-0.84ex}E_x\,$}
\newcommand{\MEy}{$\not\hspace*{-0.84ex}E_y\,$}

\newcommand{\size}     {0.25}
\newcommand{\siz}      {0.4}

\newcommand{\ttb}     {$t\overline{t}$ }
\newcommand{\wbb}     {$Wb\overline{b}$ }
\newcommand{\st}      {Single-Top }

\newcommand{\inpb}      {pb$^{-1}$}
\newcommand{\ttbar}     {$t\overline{t}$}
\newcommand{\ppbar}     {$p\overline{p}$}
\newcommand{\herwig}    {\sc{herwig}}
\newcommand{\MCFM}      {\sc{MCFM}}
\newcommand{\PYTHIA}    {\sc PYTHIA}
\newcommand{\ALPGEN}    {\sc{ALPGEN}}
\newcommand{\HERWIG}    {\sc{HERWIG}}
\newcommand{\vecbos}    {\sc{vecbos}}
\newcommand{\qq}        {\sc{qq}} \newcommand{\tauola}    {\sc{tauola}}
\newcommand{\GEANT}     {\sc{GEANT}}
\newcommand{\CTEQ}      {\sc{CTEQ5L}}
\newcommand{\COMPHEP}   {\sc{COMPHEP}}
\newcommand{\misID}     {\mbox{$\not\!\!\small{I\!D}$}}
\newcommand{\rar}       {\rightarrow}
    %%\newcommand{\MET}       {$\not\!\!E_T$ }
    %%\newcommand{\mlep}       {$\not\!\!\eel$ }

%%%%%%%%%%%%%%%%%%%%%%%%%%%%%%%%%%%

% remove the following for publication 

%\begin{figure}
%\leftline{\includegraphics[scale=0.5]{d0logo.eps}\hfill D\O notes 4946,5389}
%\end{figure}

\leftline{ } 

%\leftline{Primary authors: Gregorio Bernardi, Hyunwoo Kim,}

%\leftline{Lars Sonnenschein, Wade Fisher, for the Higgs Group } 

%\rightline {Send comments to d0-run2eb-004@fnal.gov}
%\rightline {by Nov. 27$^{th}$ 2007; to be submitted to PLB}
\rightline{FERMILAB-PUB-07/640-E}

\title{A combined search for the standard model Higgs boson at $\sqrt{s}=1.96$~TeV} 
%\input list_of_authors_r2.tex
%\author{                                                                      
% LIST_OF_AUTHORS_R2.TEX               10/09/07(b)          
%
\author{V.M.~Abazov$^{36}$}
\author{B.~Abbott$^{76}$}
\author{M.~Abolins$^{66}$}
\author{B.S.~Acharya$^{29}$}
\author{M.~Adams$^{52}$}
\author{T.~Adams$^{50}$}
\author{E.~Aguilo$^{6}$}
\author{S.H.~Ahn$^{31}$}
\author{M.~Ahsan$^{60}$}
\author{G.D.~Alexeev$^{36}$}
% ...

\affiliation{\vspace{0.1 in}(The D\O\ Collaboration)\vspace{0.1 in}}
% ...
\affiliation{$^{6}$University of Alberta, Edmonton, Alberta, Canada,
                Simon Fraser University, Burnaby, British Columbia, Canada,
                York University, Toronto, Ontario, Canada, and
                McGill University, Montreal, Quebec, Canada}
% ...
\affiliation{$^{29}$Tata Institute of Fundamental Research, Mumbai, India}
% ...
\affiliation{$^{31}$Korea Detector Laboratory, Korea University, Seoul, Korea}
% ...
\affiliation{$^{36}$Joint Institute for Nuclear Research, Dubna, Russia}
% ...
\affiliation{$^{50}$Florida State University, Tallahassee, Florida 32306, USA}
% ...
\affiliation{$^{52}$University of Illinois at Chicago,
                Chicago, Illinois 60607, USA}
% ...
\affiliation{$^{60}$Kansas State University, Manhattan, Kansas 66506, USA}
% ...
\affiliation{$^{66}$Michigan State University,
                East Lansing, Michigan 48824, USA}
% ...
\affiliation{$^{76}$University of Oklahoma, Norman, Oklahoma 73019, USA}
% ...

\date{December 4, 2007}

\begin{abstract}
%\begin{center}
%\begin{minipage}{.8\textwidth}
%{\small 
We present new results of the search for $WH \rightarrow \ell \nu b \bar{b}$ production
in $p\bar{p}$ collisions 
at a center of mass energy of $\sqrt{s}=1.96\,\mbox{TeV}$, based on a dataset with integrated
luminosity of 0.44 fb$^{-1}$. We  combine these new results with previously published searches by
the D0 collaboration, %  with a similar luminosity,
for $WH$ and $ZH$ production analyzed in the \MET $b \bar{b}$ final state, 
    %% on $WH$ ($\rightarrow \mlep \nu \bar{b}$ production,
    %%on $ZH$ ($\rig htarrow \nu \bar{\nu} b \bar{b}$)
for $ZH$ ($\rightarrow \ell^+\ell^- b \bar{b}$) 
production, for $WH (\rightarrow WWW^{}$) production, and for $H$ ($\rightarrow WW^{}$) direct
production.
No signal-like excess is observed either in the $WH$ analysis or in the combination 
of all D0 Higgs boson 
analyses. We set 95\% C.L. (expected)
upper limits on 
%$WH$ production cross section times $B$(\hbb) ranging from 1.6 (2.2)~pb
$\sigma(p\bar{p} \rightarrow WH) \times B$(\hbb) ranging from 1.6 (2.2)~pb
to 1.9 (3.3)~pb for Higgs boson masses between 105 
and 145~GeV, 
to be compared to the theoretical prediction of 
0.13 pb for a standard model (SM) Higgs boson with mass $m_H=115$ GeV.
After combination with the other D0 Higgs boson searches, we
obtain for $m_H=115$ GeV an observed (expected) limit  8.5 (12.1)  times higher than the SM 
predicted
Higgs boson production cross section. For $m_H=160$ GeV, the corresponding observed (expected)
ratio is 10.2 (9.0).
\end{abstract}
\maketitle

Spontaneous electroweak symmetry breaking in the standard model (SM) provides
an explanation for the masses of the elementary particles, otherwise massless
in the unbroken gauge theory. Its success, in particular in explaining the
mass of the electroweak vector bosons, awaits one last but necessary
experimental confirmation: the observation of the Higgs boson, which is a 
scalar particle associated with the symmetry breaking. 
For Higgs boson searches, the most sensitive  production channel at the Tevatron 
for a Higgs boson with mass below 
130 GeV is the associated production of a Higgs boson with a $W$ boson.
All possible channels, however,  must be studied to gain sensitivity through their combination.

At a center-of-mass energy of $\sqrt{s}=1.96\,\mbox{TeV}$, three $p \bar{p} \rightarrow WH$ 
searches have already been  published or submitted for publication, one~\cite{emu-hep-ex/0410062} 
using a subsample (0.17~fb$^{-1}$) 
of the dataset used in this letter, while
the two others are from the CDF collaboration: one uses
0.32~fb$^{-1}$~\cite{emu-CDF-wh} of data, the other updates it using improved analysis 
techniques and a larger dataset based on
1.0~fb$^{-1}$
of integrated luminosity~\cite{emu-CDF-wh-1fb}.

For this  $WH$ analysis
we require one high transverse momentum ($p_T$) lepton ($e$ or $\mu$), missing transverse energy \MET$\rm{}$ 
to account for the neutrino in the $W$ boson decay, and 
exactly two jets
with at least one of them being identified as originating from a bottom ($b$) quark 
jet (``$b$-tagged''),
as detailed below.
The dominant backgrounds to $WH$ production are  $W+$ heavy-flavor production, 
top quark pair production ($t\bar{t}$), and
single top quark production. 
This  analysis uses a dataset of 0.44~fb$^{-1}$. 
Compared to
the previous D0 result, 
 the $b$-jet identification
has been optimized, and the muon channel 
has been added.

    %%It has been shown \cite{emu-hep-ex/041006}
    %%that the signal to background ratio can be improved by requiring exactly
    %%two jets, so we concentrate on this signature in this analysis.
The result of this search is then combined
with previously published searches by
the D0 collaboration  with a similar luminosity. These searches cover
$WH$ and $ZH$ production analyzed in the \MET $b \bar{b}$ final state~\cite{emu-dzZHv},
    %% on $WH$ ($\rightarrow \mlep \nu \bar{b}$ production,
    %%on $ZH$ ($\rightarrow \nu \bar{\nu} b \bar{b}$)
$ZH$ ($\rightarrow \ell^+\ell^- b \bar{b}$)
production~\cite{emu-dzZHl}, $WH (\rightarrow WW^+W^{-}$)
production~\cite{emu-dzWWW},  and $H$ ($\rightarrow W^+W^{-}$) direct
production~\cite{emu-dzHWW}. %Note that some of the $W$'s can be virtual.
In the following, the particle charges will not be mentioned explicitly, except 
when needed to resolve potential ambiguity.
We 
first describe the $WH$ analysis in detail, then the full combination of results.

    %%--------------------------------------------------------
    %%\section{Data Sample and Event Selection}
    %%--------------------------------------------------------

The  $WH$ analysis relies on the following 
components of the D0 detector~\cite{emu-run1det,emu-run2det}:\\
{\it i)} a central-tracking system, which consists of a silicon microstrip tracker (SMT)
 and a central fiber tracker, both located within a 2~T
superconducting solenoidal magnet;\\
{\it ii)} a  liquid-argon/uranium calorimeter with a central 
section (CC) covering pseudorapidity~\cite{foot1}
%{The pseudorapidity is defined as a function
%of the polar angle $\theta$ as $\eta \equiv - \ln(\tan{\frac{\theta}{2}})$}
 $|\eta|$  $< 1.1$, and two end
calorimeters (EC) extending coverage to $|\eta| \simeq 3.2$, all housed in 
separate cryostats,
and with
scintillators between the CC and EC cryostats providing 
sampling of developing showers at $1.1<|\eta|<1.4$;\\
{\it iii)} a muon system, which  surrounds the calorimeter and consists of a 
layer of tracking detectors and scintillation trigger counters 
before 1.8~T iron toroids, followed by two more similar layers behind
the toroids. 
%The muon tracking at $|\eta|<1$ relies on 10~cm wide drift
%tubes~\cite{emu-run1det}, while 1~cm mini-drift tubes are used at
%$1<|\eta|<2$.

    %%The trigger and data acquisition systems are designed to accommodate 
    %%the large luminosity of Run II. 
We reject  data periods in which the quality of the data in the tracking, 
the calorimeter, or the muon system is compromised.
The luminosity is measured using plastic scintillator arrays located in front 
of the EC cryostats, covering $2.7 < |\eta| < 4.4$. 
The uncertainty on the
measured luminosity is ~6.1\%. %~\cite{emu-lumi}.
    %%This data rejection is made at different levels, and is  
    %%run-based or luminosity-block based (a period which lasts about 1-2 minutes).
    %%The resulting luminosity  (170 \inpb) is about 10\% lower than the total
    %%recorded luminosity during this period.
The $W$ + jets candidate events
must pass one of the triggers which require, for the $e$ channel, at 
least one electromagnetic (EM) object,
and for the $\mu$ channel, at least one muon object or a  trigger requiring
a muon and a jet in the final state.

The event selection for the $WH$ analysis requires one
lepton candidate with transverse momentum $p_T > 20 $ GeV, 
\MET\ $ > 25 $ GeV, and exactly two jets with
$p_T > 20$ GeV and  $|\eta| < 2.5$.
Only events having a primary $z$ vertex within $\pm$ 60 cm of the nominal interaction 
point are accepted.
If the lepton is an electron, it is required to
have  $| \eta | <  1.1$. If it is a muon the requirement is
 $| \eta | <  2.0$.
    %%%%%%%%to have  $| \eta^{\ell}_{detector} | <  1.1$. If it is a muon the requirement is
    %%%%%%%% $| \eta^{\ell}_{detector} | <  2.0$

Electrons are identified in two steps.
The preselected electron candidates (seeded by an energy cluster in the EM calorimeter) are first required to 
satisfy
identification (ID) criteria: (a) a large fraction of their energy deposited
in EM layers, i.e. {\rm EMF} $> 0.9$, (b) low fractional 
energy deposited around the expected electron energy deposition,
%isolation $<0.10$, 
and  (c)
spatial energy distribution in the EM calorimeter
consistent with that of an electron.
% shower shape requirements.
These criteria define  ``loose'' electrons.
The loose electrons are then tested with a
likelihood  algorithm, optimized on $Z \rightarrow ee$  samples,
and which takes as input seven quantities sensitive to the EM nature 
of the particles~\cite{ttbar-prd}. If they satisfy the likelihood requirement, they are accepted
as final (``tight'') electrons for the analysis.
The efficiencies of the ID
and likelihood requirements 
are determined from a dielectron sample in which
we select a pure set of $Z$ events. 
The combined reconstruction and ID efficiency is found to be
$(95.4 \pm 0.4)\%$.
The likelihood efficiency for electrons 
is $(92.0 \pm 0.3)\%$. 

Muons are reconstructed using information from
the muon detector and the central tracker.
They are required to have hits in all 
layers of the muon system inside and outside the toroid.
The superior spatial resolution of the central tracker, inside the strong
solenoidal magnetic field, is used
to improve the accuracy of kinematic properties of the muon
and to confirm that the muon originated from the primary vertex.
A veto against cosmic-ray muons based on the timing of hits in the muon-system scintillator detectors 
is applied.
Quality criteria on the associated central track are also applied
to reject the majority of background  muons: a small
track impact parameter ($dca$) compared to its                                                                       
resolution ($\sigma_{{{dca}}}$) is required,                                                                          
$dca < 3 \sigma_{{\text {dca}}}$, 
to reject  muons originating from semi-leptonic decays of heavy-flavor
hadrons which constitute the main background. Such background muons
have a lower transverse momentum spectrum and are not typically isolated
due to jet fragmentation.
A loose isolation criterion is defined using the spatial separation
 $\Delta R =\sqrt{(\Delta \eta)^2 + (\Delta \varphi)^2}$
between a muon and the closest jet in the 
 $\eta$--$\varphi$ plane, where $\varphi$ is the azimuthal angle,
    %%The distance
    %%between two objects in this plane can be defined by
    %%$\Delta R =\sqrt{\Delta \eta^2 + \Delta \varphi^2}$. Here,
we require 
    %%the distance  between a muon and the closest jet  to 
    %%be 
$\Delta R  >0.5$.
Tighter muon isolation criteria are defined by requiring that
    %%\begin{itemize}
    %%a) Halo(0.1, 0.4)$<$ 2.5~GeV, where Halo(0.1, 0.4) is 
the scalar sum of the transverse energy of
 calorimeter clusters in a hollow cone 
($0.1 <\Delta R <0.4$) around the muon divided by the $p_T$
of the muon  be less than 0.08, and
    %%Only electromagnetic
    %%and fine hadronic calorimeter cells are taken into account in the calculation
    %%of this variable, since there is significantly
    %%more noise in the coarse hadronic calorimeter.
    %%b) TrkCone$(0.5)<2.5\,$GeV, where TrkCone(0.5) is 
 the scalar sum of
the transverse momenta of all tracks within a cone of radius $\Delta R=0.5$
around the muon divided by the $p_T$
of the muon  be less than 0.06.
The track matched to the muon is excluded from  this sum.

The jets are reconstructed using
 a cone algorithm~\cite{blazey} with a radius
of $\Delta R=0.5$. We apply standard D0 jet-ID criteria to avoid fake jets which 
occasionally originate from noise in the calorimeter,
    %% identified following the CALGO/Jet-id recommendations \cite{emu-jetmet}. 
    %%The jet multiplicities and properties may be
    %%distorted by the presence of so-called ``noise'' jets which are reduced by
    %%the jet-ID criteria.  Noise jets are either created by fake energy in
    %%the calorimeter (electronics malfunction in some region of the
    %%calorimeter), or they are real low energy jets which collect sufficient
    %%fake energy to pass the energy threshold requirements.  Since distinguishing
    %%etween these two cases on an event by event basis is not
    %%straightforward, the CALGO/Jet-ID group has developed a set of ID requirements which
    %%classify the jets as ``good'' or ``bad''.  The energy of the ``bad''
    %%jets is not corrected and this decreases their influence on the event kinematics. 
    %%A subset of the bad jets consists of noise jets,  
    %%i.e. jets which fail the L1 criteria \cite{emu-d0-subskim}.
    %%From here on jets will always refer to ``good jets''.
i.e., the energy fraction in the EM layers of a jet is required to be
      $ 0.05 < $ EMF $< 0.95 $ and the
energy fraction in the CH section
 of the calorimeter is required to be $< 0.4$.
    %% The effect of noise jets is strongly
    %% diminished by 
    %% the use of the T42 algorithm %~\cite{emu-T42-2}
    %%  which removes on average $15\,$GeV of noise per event~\cite{emu-T42-3}.
    %% %\end{itemize}
The difference in efficiency of the jet-ID requirements
between data and simulation is quantified
in the overall jet reconstruction efficiency scale factor
to which a systematic uncertainty of $5\%$ (per jet) is assigned.

The  multijet  background is estimated from the loose and tight 
$e$ or $\mu$ final 
samples.
%for every differential distribution, bin-by-bin,
as described in Ref.~\cite{ttbar-prd} using the following probabilities.
We determine from the data
the probability $p^{\text{multijet}}_{{\text{loose}} \rightarrow {\text{tight}}}$ 
 for a ``loose'' lepton originating from a jet
to pass the tight lepton requirements. This is done separately for the electron and the
muon channel and this probability is determined as a function of the
$p_T$ of the candidate lepton.
The sample of
multijet events containing a loose
lepton is selected with kinematic criteria
that ensure negligible contamination of
real leptons. 
We also determine the same type of probability 
$p^{\text{signal}}_{{\text{loose}} \rightarrow {\text{tight}}}$ 
%$p^{signal}_{loose \rightarrow tight}$ 
for a genuine isolated lepton
from $Z \rightarrow \ell^+\ell^-$ samples.
With these two probabilities and the numbers of loose and tight $W+2$ jet candidates,
we determine the number of multijet background events in our sample,
bin-by-bin, for every differential distribution.

\begin{figure}[b]
{
\psfig{
%figure=MuonPlots/EPS-p14-add.500-700H115/Graph_8_1.eps,width=3.in,clip=}
figure=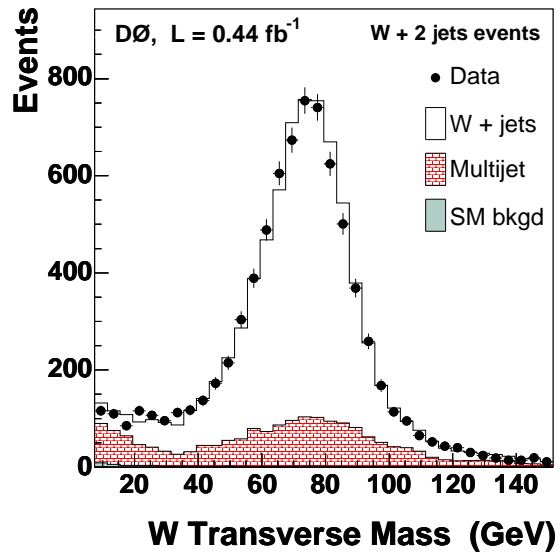,width=3.in,clip=}
}
\caption{
Distributions of 
the  transverse $W$ boson mass 
compared 
to the simulated expectation in the $W +2$ jet event sample.
The simulation is normalized to the integrated luminosity of the   data sample using the expected cross sections 
%(absolute normalization) except for the $W+$ jets sample which is normalized  to the data, 
taking into account all the other backgrounds (the fraction of $WH$ events 
is negligible before $b$-tagging).
}
\label{emu-scalar-e-mu}
\end{figure}
To select $W$ boson decays,
we require  \MET$> 25$~GeV.
The  \MET\ is calculated from the calorimeter
cells except for
unclustered cells in the outermost layer 
of the calorimeter (coarse hadronic layer, CH) %~\cite{emu-T42-1},
and is corrected when one or several
muons are present. All energy corrections to electrons or  
jets are also propagated into the \MET . 
 The transverse mass of the $W$ boson candidates in the $W + $ jets sample is
 reconstructed from
 the lepton and missing transverse energies. Its distribution
 is shown in Fig.~\ref{emu-scalar-e-mu}
 and compared 
with the sum of contributions
from multijet
events with misidentified leptons and from SM processes which are
obtained from simulated events.

The following processes
%which are used to compare with data,
are simulated  with
the  {\sc pythia} \cite{emu-pythia} MC event generator version 6.202,
making use of the CTEQ5L ~\cite{emu-CTEQ} leading-order 
parton distribution functions:
inclusive production of $W \rar e/\mu/\tau +\nu$;
 $Z \rar e e/\mu\mu/\tau\tau$;\
    %% $W \rar \tau \nu$;
    %%$Wjj \rar e/\mu/\tau +\nu jj$;
$WW,~WZ,~ZZ$; 
$t \bar{t} \rar e/\mu/\tau $ + jets production (lepton+jets and dilepton channels),
$WH \rar e/\mu/\tau +\nu + b \bar{b}$ production.
The single top quark %($s$-channel($tb$) and $t$-channel ($tqb$)) 
processes are generated using {\sc comphep} ~\cite{emu-COMPHEP}.

Throughout this Letter, ``$W$+jets'' simulated events refer to events
with a $W$ produced in association with light-flavor jets (originating
from $u$, $d$, $s$ quarks or gluons; generically denoted by $j$)
or charm jets (originating from a $c$ quark).
They constitute the dominant background before $b$-tagging.
and are generated with {\sc alpgen}~\cite{emu-ALPGEN} (interfaced
to {\sc pythia}\ for showering and fragmentation), since {\sc alpgen}\
has a more complete simulation of processes with high jet multiplicities.
The generation is based on $W+$ 2 jets ($Wjj$)  processes, including the
charm quark ($c$) processes $W c \bar{c}$ and $W c j$.
%, but not
%$W b \bar{b}$ which is 
%generated separately. 
The $W b \bar{b}$ events are  generated separately %with  {\sc ALPGEN}\  %matched with {\sc pythia}\ 
requiring two $b$ parton jets with $p_T > 8$ GeV
separated  by $\Delta R > 0.4$; its
NLO cross section is obtained using {\sc mcfm}~\cite{emu-mcfm}.

These simulated backgrounds are absolutely normalized
 (according to NLO cross sections)
%i.e. according to NLO cross sections,
 with the exception of the $W+$ jets sample
which is normalized to the data after subtraction of all the other backgrounds.
The systematic uncertainty on the NLO cross sections of these processes
is 6--18\%, depending on the process.
All these events are processed through the D0 
detector simulation, %(D\O gstar)
based on {\sc geant} \cite{geant}, %the electronics simulation 
%(D\O sim) 
and the reconstruction software. 
%(D\O reco). 
The simulated events are then weighted by the trigger efficiency
and by the data/simulation ratio of all the selection efficiencies.
The shape  of the distribution  of the
transverse mass of the $W$ candidates (Fig.~\ref{emu-scalar-e-mu})
is well reproduced
by the  simulation of the $W$ + jets processes, 
after adding the multijet background and the other SM
backgrounds.

To identify heavy-flavor jets we use a $b$-tagging algorithm 
which computes a probability  correlated to the $b$ quark lifetime~\cite{emu-JLIP05}.
%correspond approximately to a mistag rate (tagging of light flavor jets) 
%of the same value than the jet lifetime probability.
%i.e. 1\% and 0.1\%.
The requirements on the ``jet lifetime probability'' (JLIP)  have been optimized for
events with one or two $b$-jet candidates
by maximizing the sensitivity to
the Higgs boson signal.
The requirement is first set to 1\%; if two jets are tagged
the event is selected as double $b$-tagged (DT). Otherwise the requirement is tightened to 0.1\% 
and if exactly one jet is tagged the event is selected as single $b$-tagged (ST).
In this way the single and double $b$-tagged 
subsamples are independent, which  simplifies
their combination. 
The  mistag rate (tagging of light flavor jets) obtained 
in these samples are 
approximately equal to the corresponding JLIP  requirements, while
the  efficiency for correctly identifying a genuine $b$ jet 
(``$b$-tagging efficiency'') is
$(55\pm 4)\%$  and  $(33\pm 4)\%$,
respectively. 
    %%55pm4%->7.3% rel., 33pm4%->12.1% rel.
These efficiencies were determined with central ``taggable''
 jets ($|\eta|<1.2$) having a transverse momentum
of $35<p_{T}<55\,\mbox{GeV}$.
A jet is ``taggable'' if at least 2 tracks (one with $p_T>1$ GeV, the other with 
$p_T>0.5$ GeV)
 and                                             
  $\geq 1$ SMT hits  are inside the $\Delta R <0.5$ cone defining the jet.
The jet taggability is typically 80\% in a two-jet 
sample with an uncertainty of 3\%.

%In this analysis, 
For each tagged jet in the simulation, 
we apply the ratio between the expected taggability times
$b$-tagging efficiency in data and in simulation to reweight 
the simulated events. 
    %% in which one or more jets %which originate from a $b$ quark 
    %%are tagged. 
For the tagging efficiency of simulated $b$ or $c$ jets, we use $p_T-\eta$
dependent data vs. simulation scale factors, determined 
from real $b$ jets~\cite{emu-JLIP05}.
In the simulation, the tagged light flavor jets are weighted to                                                    
reproduce the mistag rate as measured in data using dedicated                                                      
samples~\cite{emu-JLIP05}.
%In simulation, the mistagged 
%light quark jets  are reweighted to match 
%the yield of such  events expected 
%in our simulated ``$W$+light jets'' sample
%using light jet tag rate functions which are determined independently using dedicated
%data samples~\cite{emu-JLIP05}.
%
%
% A $\Delta R$ requirement between the two leading 
% jets is applied ($\Delta R >0.75$) in order to reduce the influence
% of $b$ jets induced by gluon splitting and to allow an unambiguous assignment
% in the simulation of the jet flavor. The flavor of a simulated 
% jet is defined as the
% ``heaviest'' flavor found in an $\eta-\varphi$ cone with 0.3 radius, 
% centered around the direction of the reconstructed jet.
    %%, the flavors
    %%being ranked from lowest to highest as $u,d,s,c,b$.

\begin{figure}[b]
 \psfig{figure=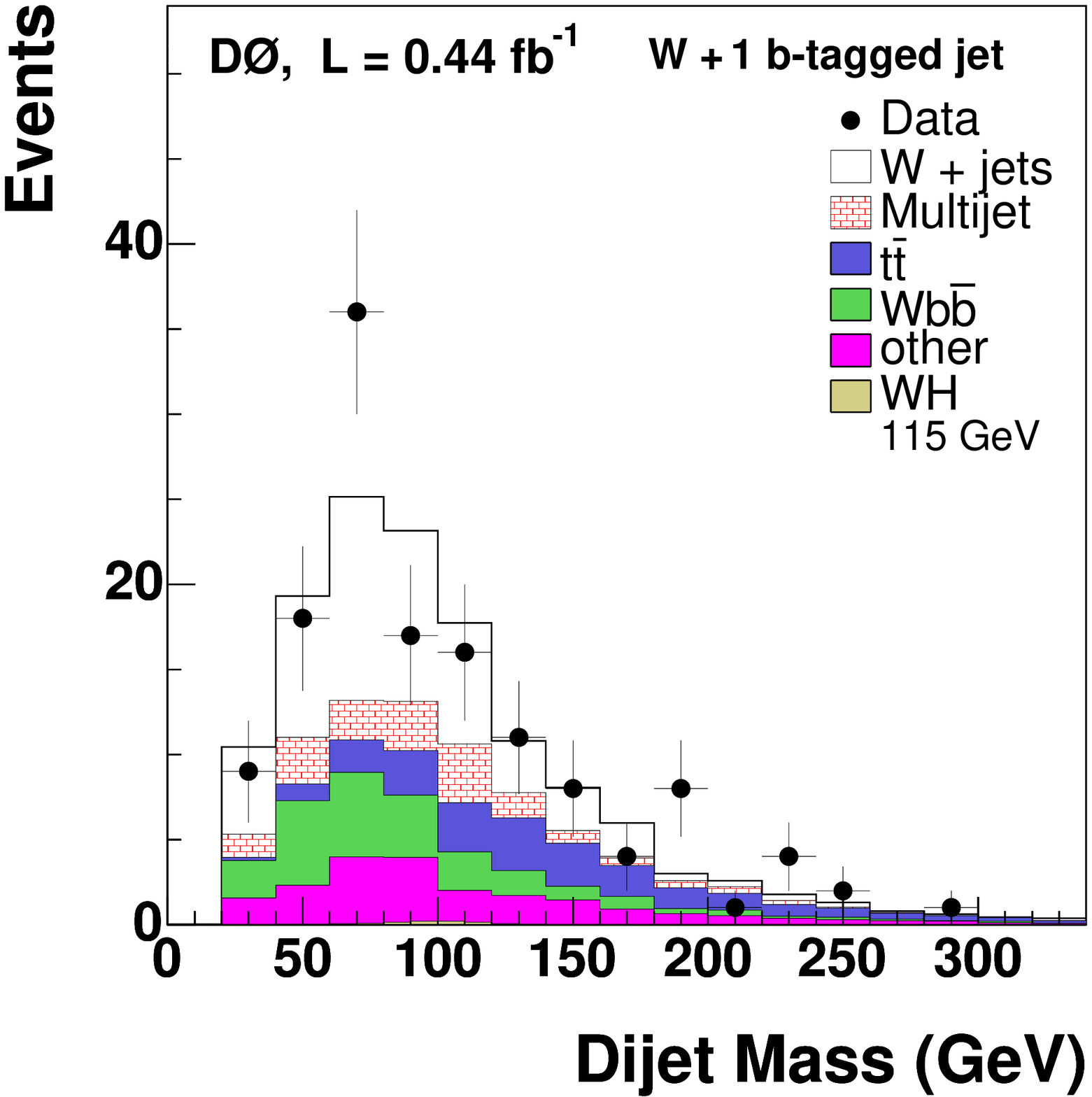,height=7.cm}\\
 \psfig{figure=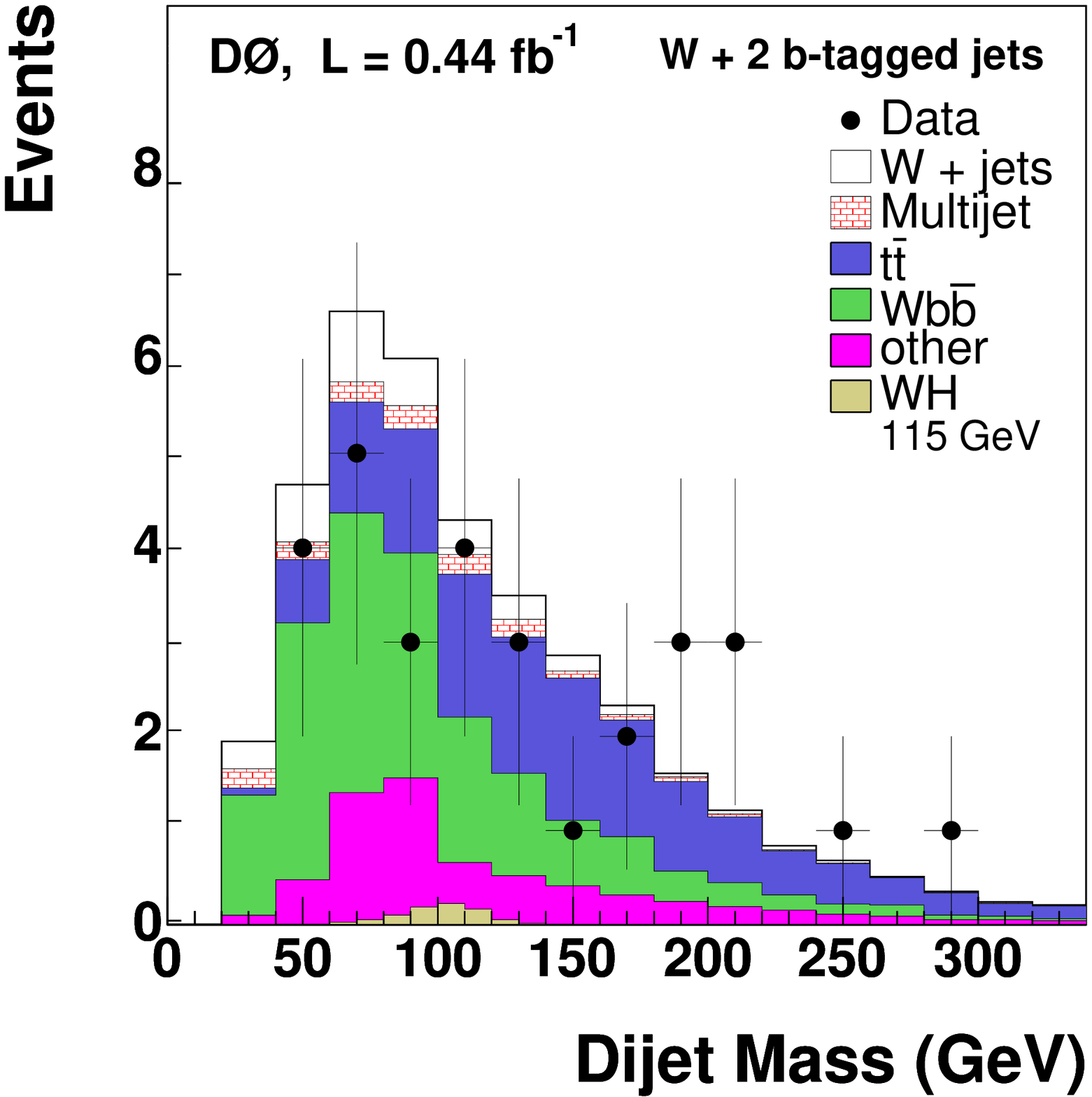,height=7.cm}
%figure =MuonPlots/EPS-p14-add.500-700H115/Graph_44_1.eps,height=7.cm}\\
%\psfig{
%figure =MuonPlots/EPS-p14-add.500-700H115/Graph_47_1.eps,height=7.cm}
 \begin{picture}(0,0)(0,0)
  \put (-160,365){\scriptsize {\bf (a)}}
  \put (-160,163){\scriptsize {\bf (b)}}
 \end{picture}
\caption{\small
Dijet mass distributions for the $W+2$ jet 
events (a) when  exactly one  jet is tightly $b$-tagged
 and (b)
when the two jets are loosely $b$-tagged (see text).
The data are compared to  $W b \bar{b}$, 
$t \bar{t}$, $W$+jets and other smaller
expectations.  
The background labeled as ``other'' in the figure
is  dominated by single top quark production.
}
\label{emu-two-tags}
\end{figure}

With the above selection criteria, we observe 137 $W+2$ jet events having exactly
one  $b$-tagged jet (ST  sample)
and 30 events having both jets $b$-tagged (DT sample). 
%We first concentrate on the
%events having only one $b$-tagged jet and will discuss the double $b$-tagged
%events in the next section.
In these samples  the multijet background is estimated using as
a loose sample the $W +2$ jet ST (DT) sample in which the lepton is selected using 
the loose lepton-ID criteria.%, yielding $ 23 \pm 8$ ($1.5 \pm 0.6$)  events.
    %%In Fig.~\ref{emu-dijets}a (b) is shown the distribution of the $p_{T}$ of the
    %%$b$-tagged jets for the $W+2$ ($W\ge 3$) jet events having at least one  jet
The distribution of
 the invariant dijet mass of $W+2$ jet events for the
ST and DT samples is  shown in Fig.~\ref{emu-two-tags}a and b.
%for  when
%requiring exactly  one  jet
%tightly $b$-tagged, i.e. for the ST sample.
    %%------------------------------ combined  figure ---------------------- begin ------------
The data are compared to the sum of the simulated 
SM processes added to the multijet background.
The  agreement  indicates that
the simulation
% which includes
%the different standard model processes, 
describes the data well.
%In Fig.\ref{emu-two-tags}b shows the same distribution for the DT sample,
%also displaying good agreement between data and expectation.

The different components of the background are shown in Table~\ref{emu-tab:table3}.
The small expected contributions from a 115 GeV Higgs
are also shown, but 
no excess above the standard model backgrounds is visible in these distributions,
so we  proceed to set limits from these distributions, after systematic
uncertainty evaluation.

\begin{table}[h]
\begin{center}
    %%\begin{table*}
    %%\begin{ }
\scalebox{0.97}{
\begin{tabular}{lrclrclrcl}
    %%&\multicolumn{2}{c}{$D_{4h}^1$}&
    %%\multicolumn{2}{c}{$D_{4h}^5$}\\
\hline
\hline
                                &\ \ $W $&+&$ 2$ jet \ \ &\ \ $W $&+&$$ 2 jet\ \  &\ \ $W $&+&$ 2$ jet\ \  \\
%                                        &  pre&-&tagged            & 1 $b$&-&tagged  & 2 $b$&-&tagged \\
%    & \multicolumn{9}{|ccc|} {    pre-tagged            & 1 $b$-tagged  & 2 $b$-tagged }\\
    & \multicolumn{3}{c} { pre-tagged }& \multicolumn{3}{c} {  1 $b$-tagged  }
    & \multicolumn{3}{c} { 2 $b$-tagged }\\%         & 1 $b$-tagged  & 2 $b$-tagged }\\
\hline
$WH$                                    &  2.3  &$\pm$& 0.4  &  0.49 &$\pm$&0.07  &  0.43&$\pm$& 0.06 \\
$WW,WZ,ZZ$                              & 148.7 &$\pm$& 23.8 &  5.3  &$\pm$& 0.8  &  2.0 &$\pm$& 0.4 \\
$Wb\bar{b}$                             & 116.3 &$\pm$& 18.6 &  22.3 &$\pm$& 4.4  & 14.4 &$\pm$& 3.2 \\
$t\bar{t}$                              &  87.6 &$\pm$&  8.6 &  21.0 &$\pm$& 4.3  & 12.6 &$\pm$& 2.7 \\
Single top                              &  41.2 &$\pm$&  5.3 &  10.0 &$\pm$& 4.8  &  3.7 &$\pm$& 0.7 \\
Multijet                                &   984 &$\pm$& 153  &  22.8 &$\pm$& 7.5  &  1.5 &$\pm$& 0.6 \\
$W$/$Z$ + jets                          &  6908 &$\pm$&1076  &  57.7 &$\pm$&10.3  &  4.1 &$\pm$& 0.7 \\
\hline
Total expect.                           & 8286          &&&  139.6 &$\pm$&28.5  & 38.7 &$\pm$& 5.8 \\
Observed Ev.                            & 8286          &&&  137            &&&   30             \\
\hline
\hline
\end{tabular}
}
%%%%%                                        &\ \ $W + 2$ jet \ \ &\ \ $W +$ 2 jets\ \  &\ \ $W + 2$ jet\ \  \\
%%%%%                                       &              & (1 $b$-tagged jet)  & (2 $b$-tagged jets) \\
%%%%%\hline
%%%%%$WH$                                    &  1.68 $\pm$ 0.17 &  0.35 $\pm$0.39  &  0.31 $\pm$ 0.06 \\
%%%%%$WW, WZ, ZZ$                            & 117.8 $\pm$ 7.2  &  4.12 $\pm$0.71  &  1.62 $\pm$ 0.26 \\
%%%%%$Wb\bar{b}$                             &  87.0 $\pm$ 16.6 &  16.7 $\pm$ 3.5  & 10.83 $\pm$ 2.38 \\
%%%%%$t\bar{t}$                              &  53.7 $\pm$ 6.0  &  12.7 $\pm$ 1.9  &  7.38 $\pm$ 1.57 \\
%%%%%Single top                              &  32.7 $\pm$ 6.7  &  7.8  $\pm$ 4.1  &  2.90 $\pm$ 0.34 \\
%%%%%QCD Multijet                            &   850 $\pm$ 231  &  18.0 $\pm$ 6.3  &  1.36 $\pm$ 0.60 \\
%%%%%$W+$ or $Z$ +jets                       &  6245 $\pm$ 751  &  52.9 $\pm$ 9.4  &  4.1 $\pm$ 0.74 \\
%%%%%\hline
%%%%%Total expectation                       & 7388  $\pm$ 817 &  111.8 $\pm$17.0  & 38.39 $\pm$ 5.8 \\
%%%%%Observed Events                         & 7388            &  112             &   30             \\
%%%%%\hline
    %%\end{small}
\caption{\label{emu-tab:table3} { 
Summary table for the $\ell$ ($e$ and  $\mu$) + 2 jets +  \MET\ final state. 
Observed events in data are compared to the
expected number of  $W + 2$ jet events
before and after $b$-tagging 
in the  simulated samples of $WH$, dibosons, $Wb \bar{b}$ production,
top production ($t \bar{t}$ and single top),
multijet background, and ``$W$/$Z+$ jets'' production.
In the pre-tagged sample the $W$/$Z$ + jets contribution is normalized such
that the total expectation is normalized to the data.
    %%excluding the processes $WH$ which are counted separately.
    %%The last column shows the same comparison in the control sample of
    %%$W + \geq 3$ jets, having two $b$-tagged jets.
}
}
    %%\vspace*{-10ex}
\end{center}
\end{table}

The experimental systematic uncertainties 
on the efficiencies 
    %%(i.e. the uncertainty on the 
    %%ratio data/simulation of the efficiencies) 
and those due to to the propagation of
other systematic uncertainties (trigger, energy calibration, detector response)
which
affect the signal and SM backgrounds 
%(QCD background
%excepted, because it is derived from data, and for which the total uncertainty is
%$\sim$ 25\%) 
are the following (ranges indicate different values for the $e$ and
$\mu$ channel):
%\begin{itemize}
%\item
 (2--3)\% uncertainty from the trigger efficiency,
% derived from the data sample used in this analysis,
%\item
 (3--4)\% uncertainty for the lepton identification and reconstruction efficiency,
    %%1)\% uncertainty for muon tracking matching efficiency data vs. Monte-Carlo scale factor 
    %%\cite{emu-d0-note4350,d0-note4264}.
%\item
    %%4)\% uncertainty due to  the 2\% systematic uncertainty on the muon energy,\\
    %%1\% uncertainty for  the uncertainty due to the $\pm 2\%$ uncertainty on the added 
 (3--4)\% for the lepton energy  scale and resolution,
%\item
 5\% for the jet identification and reconstruction efficiency, % \cite{emu-JetID}. % (7.1\% for 2 jets)
%\item
%
 5\% for the  modeling uncertainty of the jet multiplicity in the simulation,
    %% data vs. simulation comparisons. 
    %% \item
    %% 4\% to take into account the effect of the small fraction ($\simeq 4\%$) of events
    %% which may not have been generated properly ({\tt alpgen + pythia} matching)
    %% 3\% for the uncertainty on modeling of the jet multiplicity in the simulation, derived from 
    %% data vs. simulation comparisons. 
    %% \item
    %% 4\% to take into account the effect of the small fraction ($\simeq 4\%$) of events
    %% which may not have been generated properly ({\tt alpgen + pythia} matching)
%\item
 (5--12)\%  due to 
     the jet energy calibration uncertainty,
     %which includes the uncertainty on the \MET\ requirement, due to 
     %the jet energy scale and to the \MET\ smearing. 
    %%\item
    %%1\% for the propagation of the uncertainty on the jet energy resolution.
%\item
 3\% for the jet taggability, and % ($4.2\%$ for 2 jets). 
%\item
 (5--6)\% for the $b$-tagging efficiency; 
 for the light quark jets these uncertainties are 
    %%an additional uncertainty, due to the rescaling procedure of the 
    %%scaling factor of 7\% (in the DT case) or 12\% (ST) is added .
9\% (DT) and 13\% (ST).
    %%of 7\% (12\%)  for double tag (single tag) i.e. for JLIP requirement value of 1.0\% (0.1\%) 
    %%respectively, is added in quadrature.
    %%This translates into an uncertainty on the total background 
    %%of 0.6\% (7.8\%) respectively.
%\end{itemize}
In summary, for $WH$ production and simulated 
backgrounds,  the experimental systematic uncertainty is (16--19)\%.
The multijet background, determined from data, has an uncertainty 
of  25\%.
The systematic uncertainty on the cross section of the simulated backgrounds
is 6--18\%, depending on the process.
The uncertainty on the luminosity is ~6.1\%. %~\cite{emu-lumi}.
%The luminosity uncertainty (6\%) is treated separately in the limit setting procedure.

The limits for $WH$ production are obtained
using 
the $CL_s$ method~\cite{emu-junkLim,emu-wadeLim}
taking the dijet invariant mass
of the $b \bar{b}$ system as the final discriminating variable.
It is performed 
 on the ST and DT
samples of the $e$ and $\mu$ channels 
independently (four analyses), which are then combined.
%Since no data excess compared to the SM backgrounds 
%is observed we have used
    %%%%the $CL_s$ method 
    %%%%using a modified frequentist approach, 
%as explained in more details below,
% In the $CL_s$ approach, the  
%binned distributions are summed over the log-likelihood ratio test
%statistic.
%%
%%
%
 The $CL_s$ approach is based on the  
%binned distributions are summed over the log-
likelihood ratio test statistic,
$ Q = L(s+b)/L(b) =\frac{e^{-(s+b)}(s+b)^n}{n!} / \frac{e^{-b} (b)^n}{n!}$,
where $s$ and $b$ are the expected numbers of signal and background 
events while $n$ is the number of data events.
For computational ease, the log-likelihood ratio $LLR(n)=-2\ln(Q)$  
is used. In order to exploit the shape information of the final    
discriminating variable, as well as combine the different channels,  
the $LLR$ values per bin and for all channels are added.             
Systematic uncertainties are incorporated into the signal and
background expectation using Gaussian sampling of individual
uncertainties.  Correlations between uncertainties across channels are handled by 
varying simultaneously 
the fluctuations of identical sources of all channels. The 95\% C.L. limits are
determined by raising the signal cross sections until the ratio of probabilities for
the signal+background hypothesis to the background-only hypothesis falls below 5\%. 

 \begin{figure}[t]
 \psfig{figure=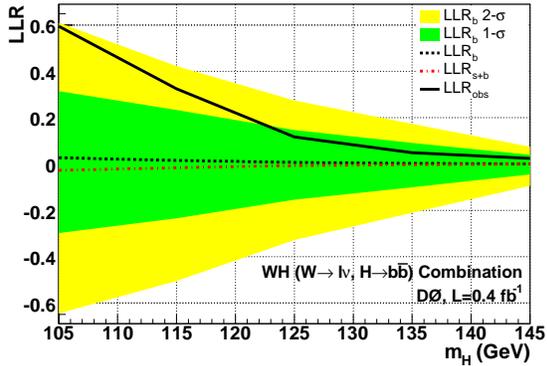,width=7.2cm}
 \caption{
 $LLR$ distributions obtained with the $CL_s$ method for the combination
of the ST and DT samples  in the $WH$ channel. 
% See D0 Note 5056 for details on $LLR$ plots.
}
 \label{emu-LLR-wh}
 \end{figure}
    %%------------------------------ combined figure ---------------------- end ------------
\begin{table}[b]
\begin{center}
%\begin{small}
%\scalebox{0.97}{ \hspace*{-2ex}
\begin{tabular}{lccccc}
\hline
\hline
$WH$ / Higgs mass [GeV]     & 105   & 115         & 125                 & 135       & 145  \\ 
%&&&&&\\
\hline
\hline
%  ST observed $\sigma$/pb&   3.25 &   3.11 &   3.16 &   3.04 &   3.68 \\
%  ST expected $\sigma$/pb&   4.50 &   4.23 &   3.94 &   3.32 &   3.62 \\
% \hline
%  DT observed $\sigma$/pb&   1.94 &   1.91 &   2.05 &   1.78 &   1.84 \\
%  DT expected $\sigma$/pb&   3.00 &   2.62 &   2.47 &   2.13 &   2.24 \\
% \hline
% \hline
%  ST+DT observed $\sigma$/pb &   1.39 &   1.37 &   1.52 &   1.42 &   1.52 \\
%  ST+DT expected $\sigma$/pb &   2.37 &   2.11 &   1.96 &   1.70 &   1.84 \\
% \hline
%\hline
 ST observed $\sigma \times B$   &       6.62 &   5.74 &   5.17 &   4.79 &   4.74\\
 ST expected $\sigma \times B$   &       8.11 &   6.94 &   6.10 &   4.90 &   5.08\\
\hline
 DT observed $\sigma \times B$   &       2.21 &   2.12 &   2.25 &   1.98 &   1.97\\
 DT expected $\sigma \times B$   &       3.55 &   3.07 &   2.89 &   2.43 &   2.58\\
\hline
 ST+DT observed $\sigma \times B$   &    1.92 &   1.71 &   1.79 &   1.64 &   1.77\\
 ST+DT expected $\sigma \times B$   &    3.25 &   2.83 &   2.53 &   2.16 &   2.21\\
\hline
%\hline
\hline
 ST observed ratio to SM  &        34.9 &   44.9 &   65.5 &  112.8 &  250.7\\
 ST expected ratio to SM  &        42.8 &   54.4 &   77.3 &  115.4 &  268.5\\
\hline
 DT observed ratio to SM  &        11.7 &   16.6 &   28.5 &   46.6 &  104.1\\
 DT expected ratio to SM  &        18.8 &   24.1 &   36.6 &   57.3 &  136.6\\
\hline 
 ST+DT obs. ratio to SM   &        10.1 &   13.4 &   22.6 &   38.6 &   93.4\\
 ST+DT exp. ratio to SM   &        17.1 &   22.1 &   32.0 &   51.0 &  116.7\\
\hline
\hline
\end{tabular}
%}
%\end{small}
    %%\vspace*{-3ex}
\end{center}
\caption{\label{emu-limits-wh}{ Observed and  expected  95\% C.L. limits 
on the cross section times branching fraction 
$\sigma \times B$, where $B=B$($H \rightarrow b \bar{b}$) and $\sigma$ is in pb,
for different Higgs boson mass values,
for single and  double  $b$-tagged events, and ST+DT combination in the 
$WH \rightarrow \ell \nu b \bar{b}$ 
 channel, with $\ell= e$ or $\mu$.
The corresponding ratios to the predicted SM Higgs production cross section
are also given.}}
\end{table}
Figure~\ref{emu-LLR-wh} shows the $LLR$ distributions for the $WH$
combined result. The $LLR$ values for the signal+background hypothesis
($LLR_{{s+b}}$), background-only hypothesis ($LLR_{b}$), and the
observed data ($LLR_{obs}$) are shown.
The quantities $LLR_{s+b}$, $LLR_{b}$, and
 $LLR_{obs}$ are obtained by setting $n
 = s+b,\ b$ or $n$(observed) into $LLR(n)$.
%     %%------------------------------ combined figure ---------------------- end ------------
% \begin{figure}[htbp]a
% \centerline{
% \psfig{figure=whplots/P14whlv1TagLLR.eps,width=8cm}
% {\psfig{figure=whplots/P14whlv2TagLLR.eps,width=8cm}
% }
% }
% \centerline{
% \psfig{figure=whplots/P14whlvLLR.eps,width=8cm}
% {\psfig{figure=whplots/P14whcombLLR.eps,width=8cm}
% }
% }
% \begin{picture}(0,0)(0,0)
%  \put (35,220){\scriptsize {\bf (a)}}
%  \put (265,220){\scriptsize {\bf (b)}}
%  \put (35,60 ){\scriptsize {\bf (c)}}
%  \put (265,60 ){\scriptsize {\bf (d)}}
% \end{picture}
% \caption{
% $LLR$ distributions obtained with the $CL_s$ method for the ST (a), DT (b) and ST+DT (c)
% samples  in the $WH$-leptonic channel. In (d) the $WH$-leptonic is combined with the 
% ``missing lepton'' $WH$ obtained from the Missing $E_T$+jets final state.
% See D0 Note 5056 for details on $LLR$ plots.
% }
% \label{emu-LLR-wh}
% \end{figure}
%     %%------------------------------ combined figure ---------------------- end ------------
\begin{figure}[b]
    %% {
 \centerline{
 \psfig{
   figure=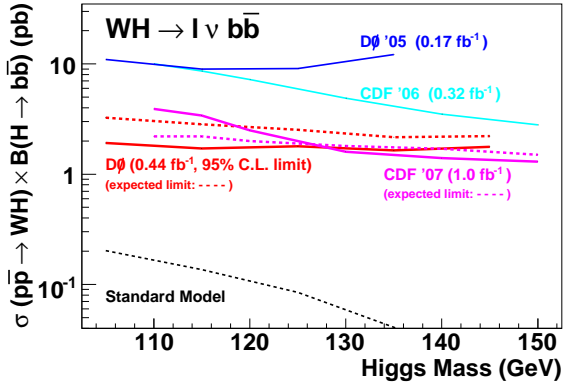,height=5.6cm}
%   figure = newlimits_WH_p14.eps,height=7.6cm}
 }
    %%\begin{picture}(0,0)(0,0)
    %% \put (145,460){\scriptsize {\bf (a)}}
    %% \put (380,460){\scriptsize {\bf (b)}}
    %% \put (145,220){\scriptsize {\bf (c)}}
    %% \put (380,220){\scriptsize {\bf (d)}}
    %%\end{picture}
\caption{ \label{emu-limits_plot}
95\% C.L.  cross section upper limit (and corresponding expected limit)
 on
$\sigma(p\bar{p} \rightarrow WH) \times B(H \rightarrow b \bar{b})$
($W$ boson decaying into a lepton + neutrino and Higgs boson into $b\bar{b}$) vs. Higgs boson mass,
compared to 
the SM expectation.
The published D0 $e$ channel observed results, based on an integrated luminosity of 
0.17~fb$^{-1}$
and the  CDF ($e+\mu$ channels) results with 0.32~fb$^{-1}$ and
1.0~fb$^{-1}$
 are also shown.
}
\end{figure}
The shaded bands represent the one  and two
standard deviation ($\sigma$) departures for $LLR_{b}$. 
These
distributions can be interpreted as follows:
The separation between $LLR_{b}$ and $LLR_{s+b}$ provides a
measure of the discriminating power of the search; 
%this is the ability of
%the analysis to separate the $s+b$ and $b-$only hypotheses.
 the width of the $LLR_{b}$ distribution 
%(shown here as one and two
%standard deviation ($\sigma$) bands) 
provides an estimate of the sensitivity of the
analysis to a signal-plus-background-like fluctuation in data, taking account of
the systematic uncertainties;
%For example, when a
%1-$\sigma$ background fluctuation is large compared to the signal
%expectation, the analysis sensitivity is thereby limited.
the value of $LLR_{obs}$ relative to $LLR_{s+b}$ and $LLR_{b}$
indicates whether the data distribution appears to be more signal-like
or background-like, and the significance of any departures
of $LLR_{obs}$ from $LLR_{b}$ can be evaluated by the width of the
$LLR_{b}$ distribution.
%\end{itemize}

The observed (expected) combined upper limits obtained at 95\% C.L.
on $\sigma(p\bar{p} \rightarrow WH) \times B(H \rightarrow b \bar{b})$
range from 1.6~pb to 1.9~pb (2.2~pb to 3.3~pb)
for Higgs boson masses between 105 and 145~GeV
and are displayed in 
%the Higgs boson cross section limit plot, 
Fig.~\ref{emu-limits_plot}. 
They are also given in Table~\ref{emu-limits-wh} 
together with the ST and DT subchannel limits and
the ratios of all these limits to the predicted SM cross section. 
%while the expected SM cross section
%for $m_H$=115 GeV is 0.13 pb 
These new $WH$ 
upper limits 
 are compared in Fig.~\ref{emu-limits_plot}
to the previously published results on $WH$ production from
D0 on 0.17~fb$^{-1}$ of data in the electron channel only~ \cite{emu-hep-ex/0410062}
and  CDF (0.32~fb$^{-1}$ 
$e$+$\mu$ channels)~\cite{emu-CDF-wh}.
The improvement in sensitivity obtained with this analysis is 
clearly visible in the region 
where the Tevatron is most sensitive to a Higgs boson with mass
in the 115-135 GeV range.
%the Tevatron has
%best sensitivity for low Higgs boson mass discovery, i.e.
% 115--135 GeV. 
The result is also compared to the CDF result recently submitted for
publication on 1.0~fb$^{-1}$ of data~\cite{emu-CDF-wh-1fb}, showing comparable
expected sensitivity when taking into account the difference in integrated luminosity.

With the limits from the $WH$ channels reported above, we
now turn to the combination of these with limits previously obtained
from other channels.
We combine our new  $WH$ results with all the other direct searches 
for SM Higgs
bosons %in \pp~collisions at \tevE\
published by D\O. 
These are searches for Higgs bosons produced in
association with vector bosons (\pzhh~\cite{emu-dzZHv,emu-dzZHl}, \pwwww~\cite{emu-dzWWW}) 
or singly through
gluon-gluon fusion (\phwww~\cite{emu-dzHWW}). The searches were conducted with data
collected during the period 2003--2005 and correspond to integrated
luminosities ranging from 0.30~\ifb~to 0.45~\ifb.
They are separated into
twelve final states (adding to the four $WH$ final states combined earlier)
and referred to as analyses in the following.  Each
analysis is designed to isolate a particular final state defined by a
Higgs boson production and decay mode. To ensure proper
combination of signals, the analyses were designed to be mutually
exclusive.% after event selections.

The sixteen analyses are
categorized by their production processes and outlined in
Table~\ref{emu-tab:chans}.  
%In the cases of $p\bar{p}\rightarrow W/ZH$ production, 
When possible, we search for both \hbb~and \hwww~decays.  
For the
\hbb~decays,
we conduct separate ST and DT
analyses, except for \zhl\ analyses where only the DT analysis has been 
performed.
% the analyses are generally separated into two orthogonal groups,
%as already described earlier for the $WH$ analysis:
%one in which two of the $b$-quarks were tagged via $b$ jet
%identification  and one
%group in which only one $b$-quark was tagged (for 
%the  \zhl~analyses, only the double-tag is
%considered so far). 
The decays of the vector bosons further define the
analyzed final states: \whe, \whm, \zhee, \zhmm, and \zhv.  There is a
sizeable amount of \WH~signal that can mimic the \zhv~final state when
the lepton is undetected, or when the lepton is a $\tau$ decaying hadronically.
 This case is treated as a separate $WH$
analysis, referred to as \lmet. 

We also include the analysis of
\wwww~final states when the associated $W$ boson and the same-charged
$W$ boson from the Higgs boson decay leptonically, thus defining six final states:
$WH\rightarrow W e^\pm \nu e^\pm \nu$, $We^\pm \nu \mu^\pm \nu$, and
$W \mu^\pm \nu \mu^\pm \nu$, which are then grouped into three
analyses:  $ e^\pm e^\pm, \mu^\pm \mu^\pm,$ and $e^\pm \mu^\pm$.
 All decays of the third $W$ boson are
included. 

In the case of \phwww~production, we again search for
leptonic $W$ boson decays with three final states, $WW\rightarrow
e \nu e \nu$, $e \nu \mu \nu$, and $\mu \nu \mu
\nu$. For the gluon-gluon fusion process, \hbb~decays are not considered
due to the large multijet background.

\begin{table}
\begin{center}
\begin{tabular}{lccc}
\hline
Channel & L & Final & Ref.\\\
       &  (\ifb)& Variable &\\\hline
\whe, ST/DT & 0.43 & Dijet mass & -- \\
\whm, ST/DT & 0.45 & Dijet mass & --  \\
\lmet, ST/DT & 0.30 & Dijet mass & \cite{emu-dzZHv} \\
\zhv,  ST/DT  & 0.30 & Dijet mass & \cite{emu-dzZHv}\\
\zhmm,  DT  & 0.37 & Dijet mass & \cite{emu-dzZHl}\\
\zhee,  DT  & 0.45 & Dijet mass & \cite{emu-dzZHl}\\
\wwww ($e^\pm e^\pm$) & 0.45 & LH discriminant & \cite{emu-dzWWW}\\
\wwww ($e^\pm \mu^\pm$) & 0.43 & LH discriminant & \cite{emu-dzWWW}\\
\wwww ($\mu^\pm \mu^\pm$) & 0.42 & LH discriminant & \cite{emu-dzWWW}\\
\hwww~($e e$)& 0.33 & $\Delta \varphi (e,e)$ & \cite{emu-dzHWW}\\
\hwww~($e \mu$)& 0.32 & $\Delta \varphi (e,\mu)$ & \cite{emu-dzHWW}\\
\hwww~($\mu \mu$)& 0.30 & $\Delta \varphi (\mu,\mu)$ & \cite{emu-dzHWW}\\
\hline
\end{tabular}
%-\end{ruledtabular}
%%}
\caption{\label{emu-tab:chans}List of analysis channels, corresponding
integrated luminosities (L), final variables for the search,
and references.
LH stands for likelihood.}
% (Ref.); the first
%two rows correspond to this letter.}
\end{center}
\end{table}
%%\section{Limit Calculations}

As before, we combine results using the $CL_s$ method.
% with a log-likelihood ratio
%(LLR) test statistic\cite{emu-cls}. 
%since it allows to
%combine individual channels taking into account systematic
%uncertainty correlations. 
Systematic uncertainties are treated as uncertainties on the
expected numbers of signal and background events, not on the outcomes of
the limit calculations. This approach ensures that the uncertainties
and their correlations are propagated to the outcome with their proper
weights. The method used here utilizes binned final-variable
distributions rather than a single-bin (fully-integrated) value.
%%\subsection{Final Variable Preparation}
In the case of the \hbb~analyses, the final variable used for limit
setting is the invariant dijet mass, as shown for the $WH$ channel
 in Fig.~\ref{emu-two-tags}.
%either when only one of the two
%jets used for the dijet mass is tagged as a $b$ jet, or when both jets
%are $b$-tagged.  
%Examples of these two types of distributions were already
%shown in in Figs~\ref{emu-two-tags}a,c.  
In the case where \hwww, the Higgs
mass cannot be directly reconstructed due to the neutrinos in the
final state. Thus, the \wwww \ analysis uses a likelihood (LH) discriminant
formed from topological variables as a final variable~\cite{emu-dzWWW}, while
%, as shown in
%Fig.~\ref{emu-fig:fvars}b, 
the \phwww~analysis uses
the separation in $\varphi$
between the final state leptons
$\Delta \varphi (\ell_1,
\ell_2)$~\cite{emu-dzHWW}.
% the difference in the angular variable
% between the two final state leptons 
%, as shown in Fig.~\ref{emu-fig:fvars}c.
Each signal and background final variable is smoothed via Gaussian
kernel estimation~\cite{emu-smooth}.  

Both signal and background
systematic uncertainties vary for the different analyses.
%The systematic uncertainties differ between analyses for both the
%signal and background. %\cite{emu-dzWHl,dzZHv,dzZHl, dzHWW,dzWWW}. 
Here
we  summarize only the largest contributions, referring to the original 
publications for details.
 All analyses carry
an uncertainty on the integrated luminosity of 6.1\%. The
\hbb~analyses have an uncertainty on the $b$-tagging rate of (5--7)\% per
tagged jet.  These analyses also have an uncertainty on the jet
energy calibration and acceptances of  8--10\%.  For the \hwww~and
\wwww~analyses, the largest experimental uncertainties are 
associated with lepton
measurement and acceptances. These values range from (3--8)\% depending
on the final state.  The largest contribution for all analyses
is the uncertainty on the background cross sections at (6--19)\%
depending on the background. The uncertainty on the expected
multijet background is dominated by the statistics of the data sample
from which it is estimated, 
hence is uncorrelated
between analyses.
%More  details
%are given in Table~\ref{emu-tab:syst}.
The systematic uncertainties for the background rates are generally
several times larger than the signal expectation itself and are thus
an important factor in the calculation of limits.  As such, each
systematic uncertainty is folded into the signal and background
expectations via Gaussian distribution. 
%These Gaussian values are
%sampled for each Poisson pseudo-experiment trial.  
Correlations
between systematic sources are carried through in the calculation.
%For example, the uncertainty on the integrated luminosity is held to
%be correlated between all signals and backgrounds and, thus, the same
%fluctuation in the luminosity is common to all channels for a single
%MC trial. 
All systematic uncertainties originating from a common
source, see Table~\ref{emu-tab:syst}, are taken to be correlated.

To minimize the effect of systematic uncertainties on the search sensitivity, 
the individual background contributions are fitted to the data observation by minimizing a 
profile likelihood function~\cite{emu-wadeLim}. The fit computes the optimal central
values for the systematic uncertainties, while accounting for departures from the nominal 
predictions by including
a term in the $\chi^2$ function which sums the squared deviation of each systematic
uncertainty 
in units normalized by its $\pm 1 \sigma$ 
uncertainties. 
A fit is performed to the background-only hypothesis 
%separately for each trial 
and is
constrained to bins with a signal expectation smaller than 4\% of the total expected background.
%%------------------------------ combined figure ---------------------- end ------------
\begin{center}
\begin{table}[t]
%\begin{table}
\scalebox{0.90}
{
\begin{tabular}{lccc}
\hline
        &$WH, e\nu b\bar{b}$ &$WH, \mu\nu b\bar{b}$ & $WW$, \\
Source  &   DT(ST) & DT(ST)&   \small{$WWW$}\\\hline
Luminosity (\%)& 6& 6& 6\\
Jet Calibration (\%)& 4& 5 & 3 \\
Jet ID (\%)          &  7 & 7 & 0\\
Electron ID (\%)     &  7 & 0& 2\\
Muon ID (\%)         & 0 &  5 &  8 \\
$b$-tagging (\%) & 9(5)&  9(5)& 0 \\
Background $\sigma$ (\%)&6--19&6--19&6--19\\
\hline
\hline
Source & \zhv  &\zhee&\zhmm\\
       & DT(ST) & & \\\hline
Luminosity (\%)& 6& 6& 6\\
Jet Calibration (\%)& 6 &7 &7 \\
Jet ID (\%)& 7 & 7 &5 \\
Electron ID (\%)& 0& 8 &0\\
Muon ID (\%)& 0&0&12 \\
$b$-tagging (\%)& 10(7) & 12 &12 \\
Background $\sigma$ (\%)&6--19&6--19&6--19\\
\hline
\end{tabular}
 }
\caption{\label{emu-tab:syst}List of leading correlated systematic
uncertainties. The values for the systematic uncertainties are the
same for the \zhv~and \lmet~channels. Each uncertainty
is considered to be 100\% correlated across channels.
The correlated
systematic uncertainty on the background cross section ($\sigma$) is
itself subdivided according to the different background processes in
each analysis.}
%-\begin{ruledtabular}
%-\end{ruledtabular}
\end{table}
\end{center}

\vskip -26pt
To set limits on Higgs boson production ($\sigma \times B(H\rightarrow X)$) 
%via these sixteen individual
%analyses\cite{emu-dzWHl,dzZHv,dzZHl,dzHWW,dzWWW},  
the sixteen analyses
are first grouped by final state to produce individual results.  We
then group channels by production modes to form combined results and study
their respective sensitivities.
%The
%limits are derived at a confidence level (C.L.) of 95\%.  
The individual analyses  are grouped to form the
$LLR$ distributions shown in Fig.~\ref{emu-fig:LLR} for
\begin{figure}[thbp]
\psfig{figure=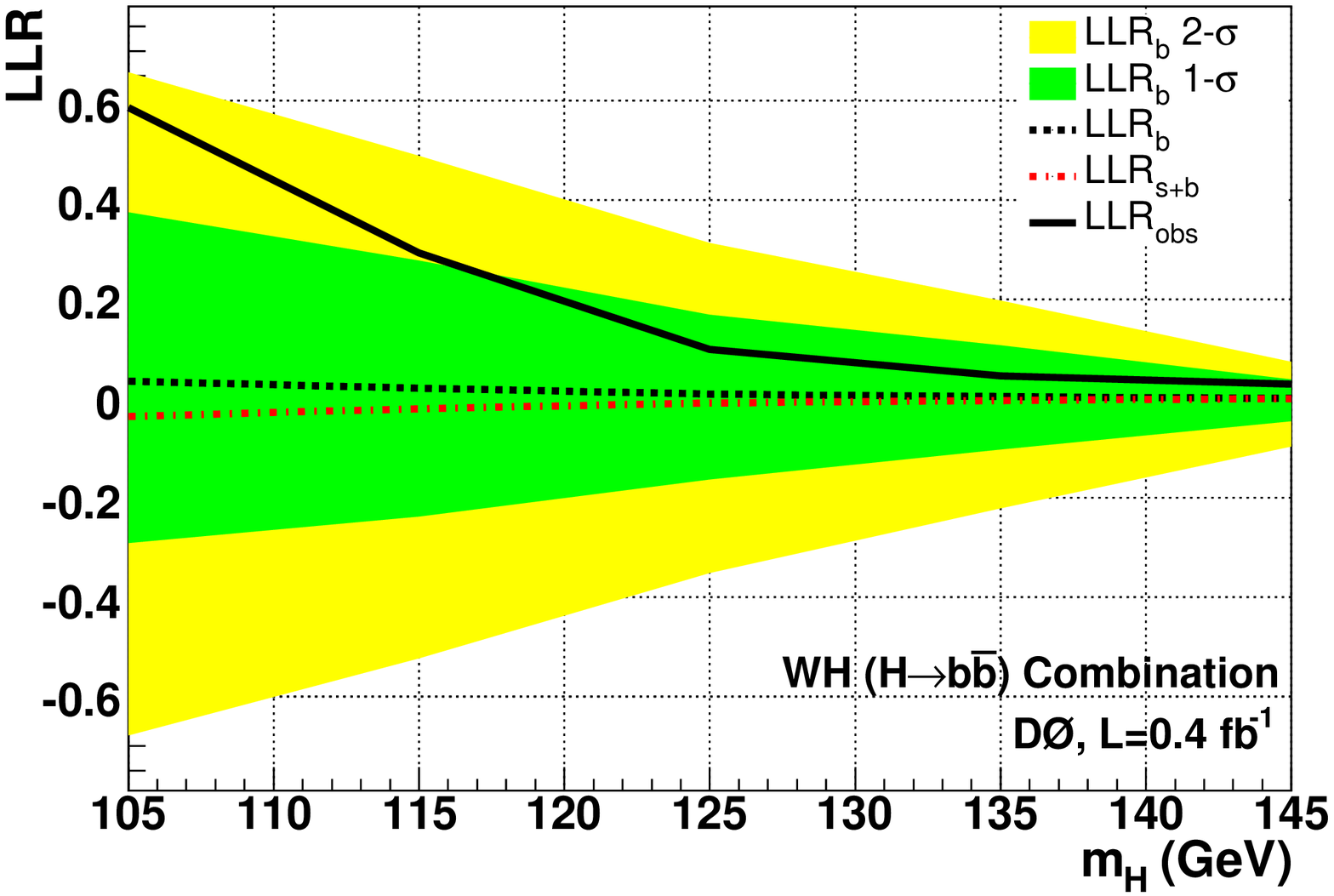,width=6.8cm}\\
\psfig{figure=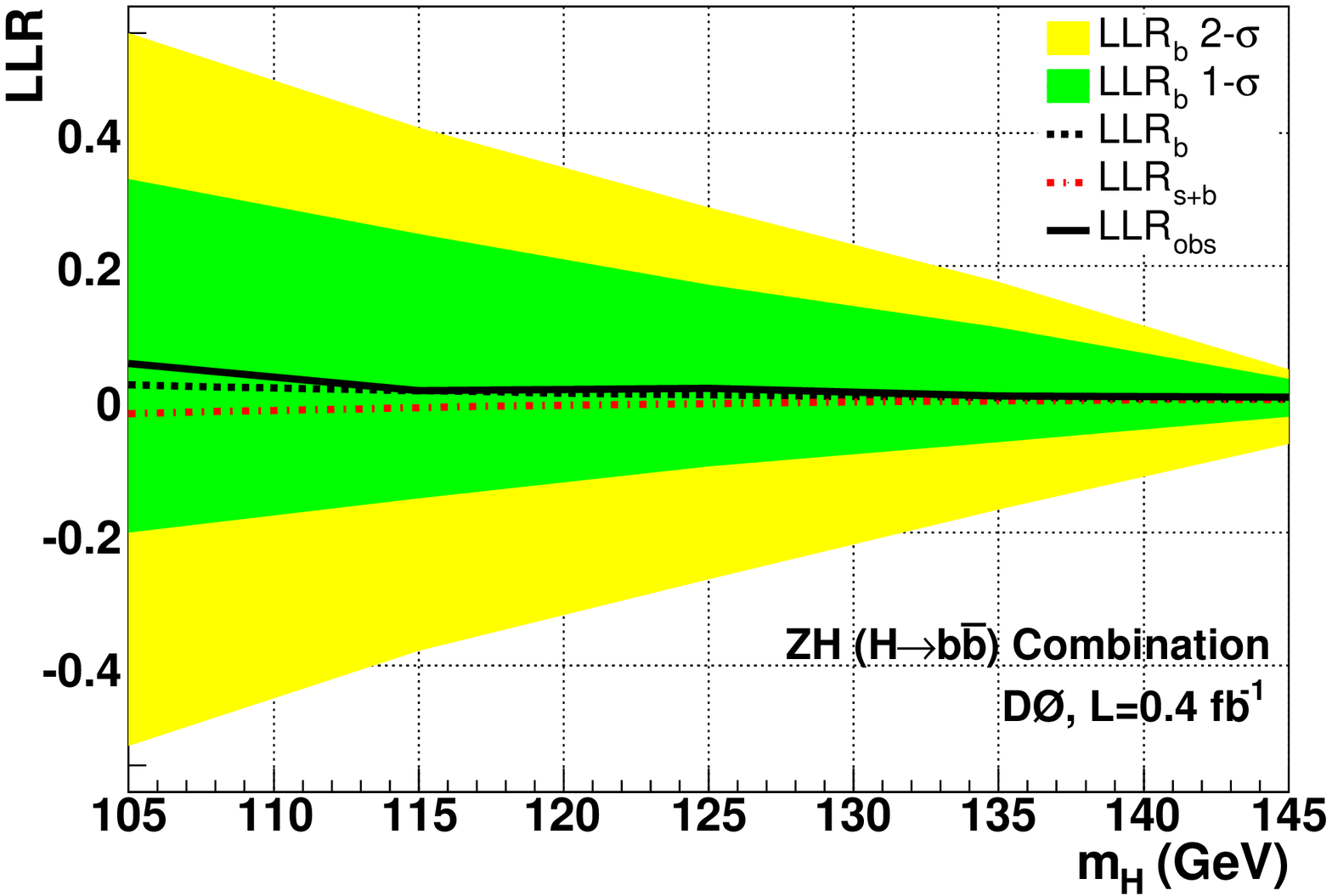,width=6.8cm}\\
\psfig{figure=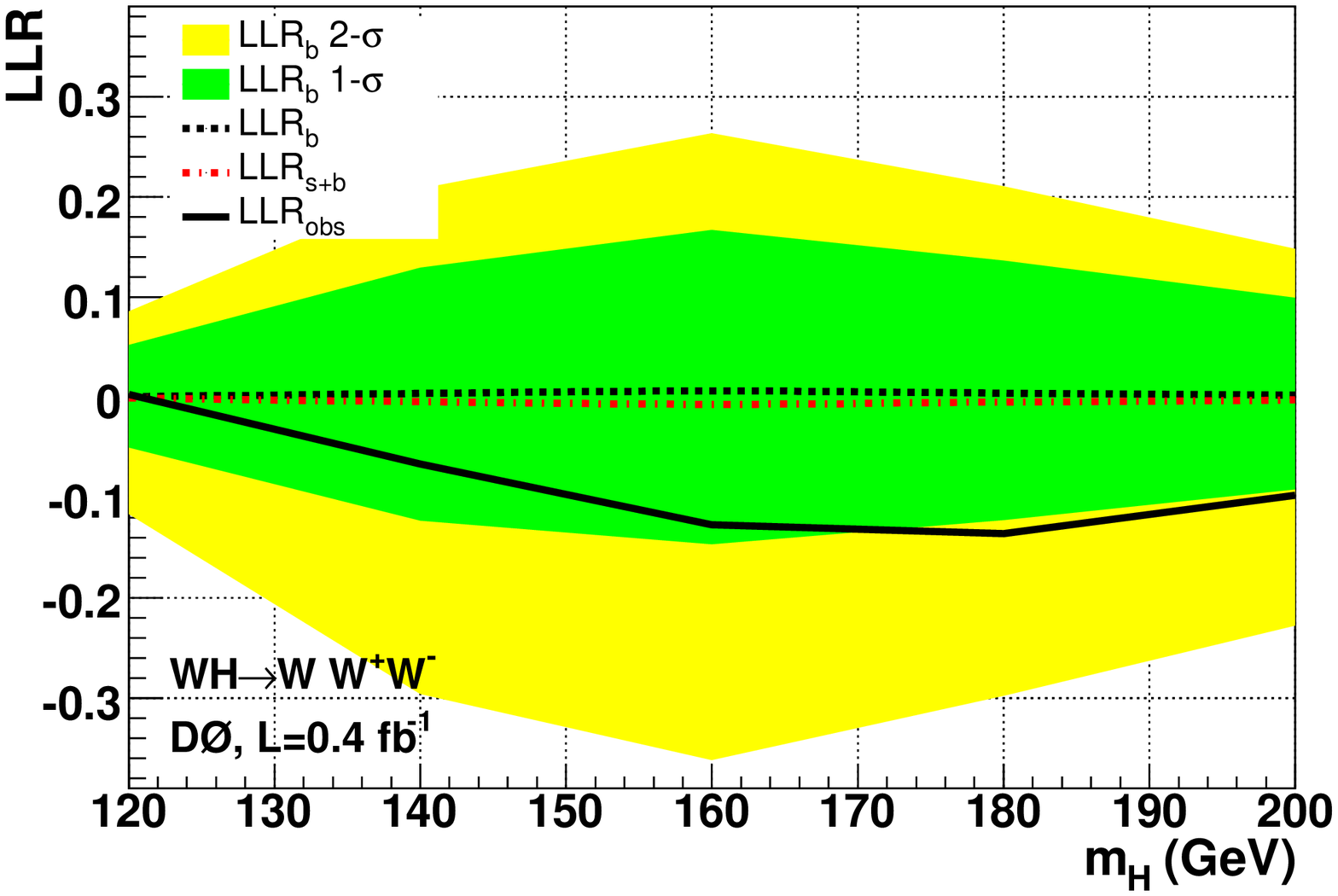,width=6.8cm}\\
\psfig{figure=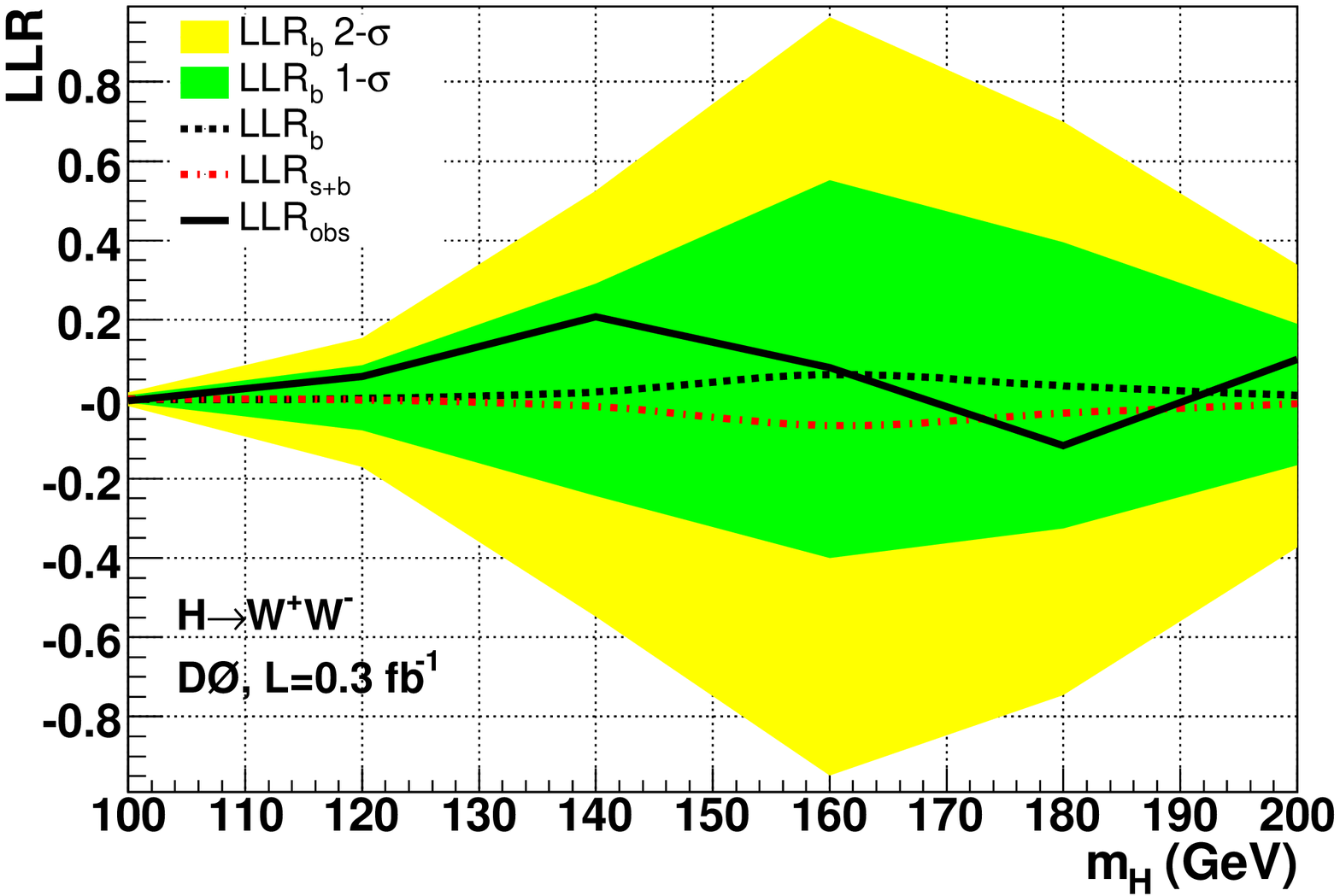,width=6.8cm}
 \begin{picture}(0,0)(0,0)
  \put (-120,420){\scriptsize {\bf (a)}}
  \put (-120,290){\scriptsize {\bf (b)}}
  \put (-120,160){\scriptsize {\bf (c)}}
  \put (-120,30){\scriptsize {\bf (d)}}
 \end{picture}
% \begin{picture}(0,0)(0,0)
%  \put (35,220){\scriptsize {\bf (a)}}
%  \put (265,220){\scriptsize {\bf (b)}}
%  \put (35,60 ){\scriptsize {\bf (c)}}
%  \put (265,60 ){\scriptsize {\bf (d)}}
% \end{picture}
\caption{
$LLR$ distributions obtained with the $CL_s$ method for the associated production
of  (a) $WH (H \rightarrow b \bar{b})$, (b) $ZH
(H \rightarrow b \bar{b}$),
(c) $WH (H \rightarrow WW)$, and (d) for the direct production channel, $H \rightarrow WW$.
See text for details.
%See  Ref.~\cite{emu-wadeLim}
%for details on $LLR$ plots.
}
\label{emu-fig:LLR}
\end{figure}
(a) all $WH$ searches, with $H \rightarrow b\bar{b}$ (ST, DT) 
in the low mass range ($m_H = 105-145$\gevc),
(b) all $ZH$ searches (ST, DT) in the same low mass range,
(c) all \wwww\ searches, over an extended mass range ($m_H = 120-200$\gevc), and
(d) all \hwww\ searches, over the full mass range ($m_H = 100-200$\gevc).
We then combine groups (a)--(d)
%group all $WH$, $ZH$, \wwww, and \hwww searches 
over the full mass range, as shown in Fig.~\ref{emu-fig:allLLR}.

%%%\vspace*{-0.4cm}
We  also compute  our results in terms of the ratio of the limits
to the SM cross section $\sigma \times B(H \rightarrow X)$ as
a function of Higgs boson mass.
%as shown in Fig.~\ref{emu-fig:allLLR}.
 The SM prediction for Higgs boson production
would therefore be excluded at 95\% C.L. when this limit
ratio falls below unity.
\begin{figure}[t]
{\centerline{
\psfig{figure=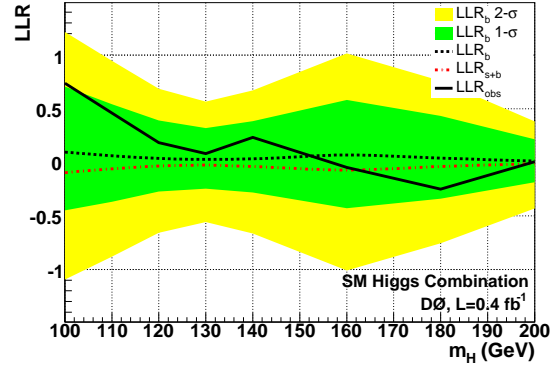,width=7.2cm}
}
}
\caption{
$LLR$ distributions obtained with the $CL_s$ method for the combination of
all channels. See text for details.
%See  Ref.~\cite{emu-wadeLim}
%for details on $LLR$ plots.
}
\label{emu-fig:allLLR}
\end{figure}
\begin{table}[b]
\begin{center}
%\begin{small}
\hspace*{-0.2cm}
\begin{tabular}{lccccc}
\hline
\hline
{Higgs mass [GeV]}                    & 105    & 115    & 125    & 135    & 145  \\ \hline
\hline
WH observed $\sigma \times B$   &      1.60 &   1.49 &   1.57 &   1.56 &   1.65 \\
WH expected $\sigma \times B$   &      2.83 &   2.38 &   2.22 &   1.89 &   2.17 \\
\hline
ZH observed $\sigma \times B$   &   2.41 &   2.23 &   1.97 &   1.77 &   3.21 \\
ZH expected $\sigma \times B$   &   2.21 &   2.02 &   1.73 &   1.52 &   2.65 \\
\hline
\hline
WH observed ratio to SM      &       8.4 &   11.7 &   19.8 &   36.7 &   87.2 \\
WH expected ratio to SM      &      14.9 &   18.6 &   28.1 &   44.5 &  114.7 \\
\hline
ZH observed ratio to SM      &   21.1 &   28.5 &   40.0 &   66.0 &  263.6 \\
ZH expected ratio to SM      &   19.4 &   25.9 &   35.2 &   56.6 &  217.4 \\
\hline
\hline
%D0 observed ratio to SM      &    5.8 &    8.5 &   13.2 &   15.5 &   11.3 \\
%D0 expected ratio to SM      &    9.6 &   12.1 &   15.5 &   15.3 &   12.8 \\
%\hline
%hline
\end{tabular}
%}
%\end{small}
    %%\vspace*{-3ex}
\end{center}
\caption{\label{emu-limits-wh-zh}{
Observed and  expected  95\% C.L. upper limits 
on the
cross section times branching fraction 
$\sigma \times B$, where $B=B(H \rightarrow b \bar{b}$),
and $\sigma$ is in pb,
for different Higgs boson mass values,
for the  $WH$ and $ZH$ combined channels
%, when $H \rightarrow b \bar{b}$
 ($WH$ includes the leptonic channels, and the
case where the charged lepton is not detected; $ZH$ includes the $ee$,$\mu\mu$, and $\nu\nu$
channels).
%The  ratios to the predicted values of the SM Higgs production cross section
%for these channels and
% for the 
%full D0 combination
%are also given.
}}
\end{table}
Table~\ref{emu-limits-wh-zh} 
shows the expected and
\begin{table*}[t]
\begin{center}
\begin{small}
\scalebox{0.97}{ \hspace*{-2ex}
\begin{tabular}{lcccccccccc}
\hline
\hline
{Higgs mass [GeV]}                                  & 100    & 110    & 115    & 120    & 130    & 140    & 160    & 180    & 200  \\
\hline                                                                         
\hline                                                                         
%WH $\rightarrow$ WWW observed $\sigma \times B$    &   --   & 71.41  & 71.41  &  11.27 &   4.41 &   1.57 &   0.09 &  0.010 &  0.004 \\
%WH $\rightarrow$ WWW expected $\sigma \times B$    &   --   & 65.35  & 65.35  &  10.78 &   3.53 &   1.30 &   0.07 &  0.007 &  0.003 \\
$WH  \rightarrow  WWW$ observed $\sigma \times B$    &   --   &   --   &   --   &  11.27 &   4.41 &   1.57 &   0.09 &  0.010 &  0.004 \\
$WH  \rightarrow  WWW$ expected $\sigma \times B$    &   --   &   --   &   --   &  10.78 &   3.53 &   1.30 &   0.07 &  0.007 &  0.003 \\
\hline                                                                         
$H  \rightarrow  WW$ observed $\sigma \times B$      &  10.79 &   5.61 &   --   &   6.07 &   5.94 &   4.24 &   3.69 &   4.07 &   3.25 \\
$H  \rightarrow  WW$ expected $\sigma \times B$      &   8.94 &   6.31 &   --   &   7.74 &   6.18 &   5.25 &   3.58 &   3.40 &   3.98 \\
\hline                                                                         
\hline                                                                         
%WH  \rightarrow  WWW$ observed ratio to SM     &   --   & 455.9  &  15.9  &  110.7 &   74.7 &   53.7 &   46.1 &   62.1 &  89.6 \\
%WH  \rightarrow  WWW$ expected ratio to SM     &   --   & 417.2  &  17.2  &  105.9 &   59.8 &   44.7 &   34.4 &   44.6 &  60.8 \\
$WH  \rightarrow  WWW$ observed ratio to SM         &   --   &   --   &   --   &  110.7 &   74.7 &   53.7 &   46.1 &   62.1 &  89.6 \\
$WH  \rightarrow  WWW$ expected ratio to SM         &   --   &   --   &   --   &  105.9 &   59.8 &   44.7 &   34.4 &   44.6 &  60.8 \\
\hline                                                                         
$H  \rightarrow  WW$ observed ratio to SM           &  636.4 &   98.9 &   --   &   46.1 &   26.4 &   14.0 &    9.9 &   15.4 &   22.2 \\
$H  \rightarrow  WW$ expected ratio to SM           &  527.5 &  111.2 &   --   &   58.8 &   27.4 &   17.3 &    9.6 &   12.8 &   27.2 \\
\hline                                                                         
\hline                                                                         
%$CL_s$ observed $\sigma \times B$                 &   2.05 &   2.20 &   2.20 &   3.16 &   4.50 &   4.15 &   3.47 &   3.80 &   3.09 \\
%$CL_s$ expected $\sigma \times B$                 &   2.71 &   2.82 &   2.82 &   3.54 &   4.68 &   5.14 &   3.20 &   3.15 &   3.67 \\
%\hline                                                                        
%$CL_s$ observed ratio to SM                  &    5.4 &    7.2 &   10.2 &   10.6 &   14.0 &   11.8 &    9.2 &   14.3 &   21.0 \\
%$CL_s$ expected ratio to SM                  &    7.1 &    9.1 &   11.1 &   11.9 &   14.6 &   14.6 &    8.5 &   11.9 &   25.0 \\
D0 observed ratio to SM                             &    5.5 &    7.1 &    8.5 &   10.5 &   14.2 &   12.8 &   10.2 &   16.1 &   23.7 \\
D0  expected ratio to SM                            &    8.7 &   10.8 &   12.1 &   14.3 &   15.7 &   13.8 &    9.0 &   12.1 &   23.5 \\
\hline
\hline
\end{tabular}
}
\end{small}
    %%\vspace*{-3ex}
\end{center}
\caption{\label{emu-limits-www-ww}{ Observed and  expected  95\% C.L. upper limits 
on the
cross section times branching fraction 
$\sigma \times B$, where $B=B(H \rightarrow WW$)
and $\sigma$ is in pb,  
for different Higgs boson mass values,
for  $WH \rightarrow WWW^{}$ and $H \rightarrow WW^{}$.
%using the $ee$ and $\mu\mu$ final states.
The  ratios to the predicted values of the SM Higgs production
cross section for these channels
and for the full D0 combination,
are also given.}}
\end{table*}
observed 95\% C.L. cross section limits and their ratios to the SM
for the $WH$ and $ZH$ analyses in the mass range $m_H = 105-145$\gevc. 
 Table~\ref{emu-limits-www-ww}   shows the same information
for  \wwww\ and \hwww\ over the full mass range.
The ratios to the SM  obtained with the full combination are also given 
and show the gain obtained by using the full information, compared to the individual channels.

The expected limits for the cross section
times branching fraction for the four groups of analyses (a)--(d) and for
the full combination, relative to the SM expectations, are shown in Fig.~\ref{emu-all-ratios}.
%%
%
%The expected values of these five combined  ratios are compared to each 
%other in Fig.
\begin{figure}[b]
{\centerline{
\psfig{figure=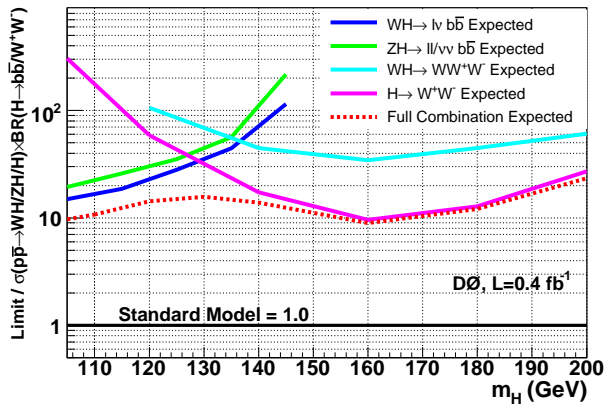,width=8cm}
}
}
\caption{
Ratios of the expected limit on the Higgs boson production cross section times
branching fraction to the SM expectation, for the different channel groups
and for
the full D0 combination.
}
\label{emu-all-ratios}
\end{figure}
For the full combination of all analyses, the expected and observed
cross section times branching ratio, relative to those for the SM, are
shown in Fig.~\ref{emu-all-data}.
%The full combination (observed and expected) is shown in  Fig.~\ref{emu-all-data}.
%
%Compared to earlier simulation-based studies  that also covered the full
%mass range~\cite{emu-shwg}, our results, which use more channels and study
%them over a wider mass range, show that the sensitivity  between
%$m_H=110-190$\gevc~is  more uniform than predicted before the beginning of the 
%Tevatron Run II. 
%
Compared to an earlier simulation
study of the Higgs boson search sensitivity conducted prior to Tevatron 
Run II~\cite{emu-shwg}, our current analyses have added new channels, 
have extended the
mass range, and show a more uniform sensitivity for 110 $< m_H <$ 190 GeV.
%Indeed
%there is only a factor of 1.5 difference in sensitivity between the most
%and the least sensitive region in this mass range.
%
%
%This suggests that the Fermilab experiments
%could be able to probe this complete mass range
%with the full Tevatron Run II luminosity.

    %%------------------------------ combined figure ---------------------- end ------------
\begin{figure}[b]
{\centerline{
\psfig{figure=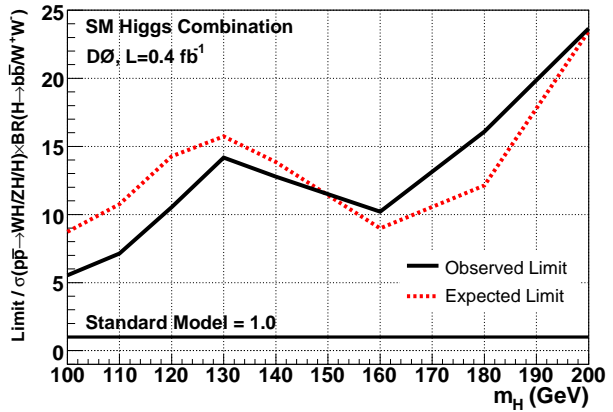,width=8cm}
}
}
\caption{
Ratios of the expected and observed limit on the Higgs boson production cross section times
branching fraction to the SM expectation, 
for the full D0 combination.
}
\label{emu-all-data}
\end{figure}
    %%------------------------------ combined figure ---------------------- end ------------

In summary,
we have presented new 95\% C.L. limits on the $WH 
\rightarrow e/\mu \nu b \bar{b}$  production cross
section times branching fraction which range  from 1.6 to 1.9 pb for 
$105 <  m_H < $ 145 GeV.  For comparison, the expected SM cross section for 
$m_H$ =115 GeV is 0.13 pb.

% we have presented new results on the $WH \rightarrow e/\mu \nu b \bar{b}$ 
% production cross section, based on 0.44 fb$^{-1}$ of 
% integrated luminosity. We set limits in this most sensitive channel between
% 1.6 and 1.9 pb for $m_H$ between 105 and 145 GeV, while the expected SM cross section
% for $m_H$=115 GeV is 0.13 pb.

We have then
combined these results with all previously published Higgs boson searches by 
the D0 collaboration
obtained with a similar luminosity (between 0.30 and 0.45 fb$^{-1}$)
to form new limits more sensitive than each
individual limit.
% and have obtained the following results:
The combined observed (expected) 95\% C.L. limit ratios to SM cross
sections for $p\bar{p} \rightarrow WH$, $H\rightarrow b\bar{b}$
range from 11.7 (18.6) at $m_H=115$\gevc~to 36.7 (44.5) at
$m_H=135$\gevc.
The combined observed (expected) 95\% C.L. limit ratios to SM cross
sections for $p\bar{p} \rightarrow ZH$, $H\rightarrow b\bar{b}$ range
from 28.5 (25.9) at $m_H=115$\gevc~to 66.0 (56.7) at $m_H=135$\gevc.
The fully combined observed (expected) 95\% C.L. limit ratio to the 
SM cross sections % on $p\bar{p} \rightarrow WH/ZH/H$, $H\rightarrow b\bar{b}/WW$ 
are 8.5 (12.1) at $m_H=115$\gevc, 10.2 (9.0) at
$m_H=160$\gevc, and 20.7 (16.0) at $m_H=190$\gevc.

These  limits and ratios will decrease in
the near future with the additional luminosity recorded at the Tevatron; more
than 2~\ifb~are currently being analyzed.
% and $\sim$8~\ifb~are expected to be delivered
%by the end of 2009.  
New techniques are being
developed 
to improve the sensitivity
through advanced multivariate techniques, neural-network $b$-tagging, and
improved di-jet mass resolution.
%
%the current sensitivity: we expect improvements via multivariate
%nalyses ($\sim 30$\% increase in sensitivity), neural-network
%$b$-tagging ($\sim 30$\%), and improved dijet mass resolution ($\sim
%20$\% for $m_H<140$\gevc). 
In addition, an anticipated combination
with the results from the CDF collaboration would yield an increase in
sensitivity of about 40\%.
With the total expected
integrated luminosity (6--8~\ifb), the Tevatron is expected to provide
sensitivity to the standard model Higgs boson beyond the current LEP limit~\cite{sm-lep}

We thank the staffs at Fermilab and collaborating institutions,
and acknowledge support from the
DOE and NSF (USA);
CEA and CNRS/IN2P3 (France);
FASI, Rosatom and RFBR (Russia);
CAPES, CNPq, FAPERJ, FAPESP and FUNDUNESP (Brazil);
DAE and DST (India);
Colciencias (Colombia);
CONACyT (Mexico);
KRF and KOSEF (Korea);
CONICET and UBACyT (Argentina);
FOM (The Netherlands);
Science and Technology Facilities Council (United Kingdom);
MSMT and GACR (Czech Republic);
CRC Program, CFI, NSERC and WestGrid Project (Canada);
BMBF and DFG (Germany);
SFI (Ireland);
The Swedish Research Council (Sweden);
CAS and CNSF (China);
Alexander von Humboldt Foundation;


\begin{thebibliography}{99}
    %%--------------------------------------------------------
% list_of_visitor_addresses_r2.tex                            10/09/07
%  available symbols are:
%  $\ast, \dag, \ddag, \S, \P, $\|$, $\ast\ast$, \dag\dag, \ddag\ddag ,\#
%
\bibitem[a]{alton}
Visitor from Augustana College, Sioux Falls, SD, USA.
\bibitem[b]{burdin}
Visitor from The University of Liverpool, Liverpool, UK.
\bibitem[c]{podesta-lerma}
Visitor from ICN-UNAM, Mexico City, Mexico.
\bibitem[d]{quadt,meyer}
Visitor from II. Physikalisches Institut, Georg-August-University G{\"o}ttingen, Germany.
\bibitem[e]{voutilainen}
Visitor from Helsinki Institute of Physics, Helsinki, Finland.
\bibitem[f]{wenger}
Visitor from Universit{\"a}t Z{\"u}rich, Z{\"u}rich, Switzerland.
%\bibitem[g]{kozminski}
%Visitor from Lewis University, Romeoville, IL, USA.

\bibitem[\dag]{IntFellows}
Fermilab International Fellow.
\bibitem[\ddag]{deceased}
Deceased.

%
\vskip 0.25cm

\bibitem{emu-hep-ex/0410062}
 D0 Collaboration, V.M.. Abazov {\it et al.},
Phys.\ Rev.\ Lett. {\bf 94}, 091802 (2005).

\bibitem{emu-CDF-wh}
CDF Collaboration, D. Acosta {\it et al.},
%{\sl Search for $H \rar b\bar{b}$
% Produced in Association with W Bosons in $p \bar{p}$ Collisions at $\sqrt{s}$ = 1.96 TeV,}
Phys.\ Rev.\ Lett. {\bf 94}, 091802 (2005).

\bibitem{emu-CDF-wh-1fb}
CDF Collaboration, T. Aaltonen {\it et al.},
%``Search for the Standard Mondel
%Higgs boson Produced in Association with $W$ bosons''
submitted to Phys.\ Rev.\ Lett.,
[arXiv:0710.4363] (2007).

    %%\bibitem{emu-foot1}
    %%When no other precision is given, ``jet'' must be understood
    %%as a jet passing standard D0 jet ID requirements, which are
    %%designed to remove rare ``noise'' jets built out
    %%of spurious energy deposits in the calorimeter.


\bibitem{emu-dzZHv} D0 Collaboration, V.M. Abazov {\it et al.}, 
%``A Search for the Standard Mondel
%Higgs boson using the \zhv~channel in \pp~Collisions at
%$\sqrt{s}=1.96$~TeV,'' submitted to Phys. Rev. Lett., [arXiv:hep-ex/0607022]
Phys.\ Rev.\ Lett.\ {\bf 97}, 161803 (2006).

\bibitem{emu-dzZHl} D0 Collaboration, V.M. Abazov {\it et al.}, 
%``A Search for \zhl~Production at  D0 in \pp~Collisions at \tevE,'' D0 Conference Note 5186
%Search for a Higgs boson produced in association with a Z boson,
%arXiv:0704.2000; Fermilab-Pub-07/076-E,
%accepted in Phys.\ Lett.\ B, 
Phys.\ Lett.\ B {\bf 655}, 209 (2007).%, 209?216


\bibitem{emu-dzWWW} D0 Collaboration, V.M. Abazov {\it et al.}, 
% ``Search for associated Higgs boson
% production $WH\rightarrow WWW^* \rightarrow \ell^\pm \nu
% \ell^{\prime\pm} \nu^\prime +X$ in $p\bar{p}$ Collisions at
% $\sqrt{s}=1.96$~TeV,'' submitted to Phys. Rev. Lett.,
% [arXiv:hep-ex/0607032]
Phys.\ Rev.\ Lett.\ {\bf 97}, 151804 (2006).

\bibitem{emu-dzHWW} D0 Collaboration, V.M. Abazov {\it et al.}, 
% ``Search for the Higgs boson in
% $H \rightarrow W W^* \rightarrow l^+ l^- (ee, e \mu)$ decays with
% 310~\ipb~at D0 in Run II,'' D0 Conference Note 5063
% \bibitem{emu-dzHWWmm} D0 Collaboration, V.M. Abazov {\it et al.}, ``Search for the Higgs boson in
% $H \rightarrow W W^* \rightarrow \mu\mu$ decays with 930~\ipb~at D0
% in Run II,'' D0 Conference Note 5194
Phys.\ Rev.\ Lett.\ {\bf 96}, 011801 (2006).

\bibitem{emu-run1det} 
 D0 Collaboration, V.M. Abazov {\it et al.},
Nucl. Instrum. and Methods A {\bf 338}, 185 (1994).

\bibitem{emu-run2det} 
D0 Collaboration, V.M. Abazov {\it et al.}, %V. Abazov {\it et al.}, 
%``The Upgraded D0 Detector'', submitted to Nucl. Instrum. Methods Phys. Res. A, [arXiv:hep-ex/0507191]
Nucl. Instr. and Methods A {\bf 565}, 463 (2006).

%D0 Collaboration, V.M. Abazov {\it et al.}, in preparation for submission to
%Nucl. Instrum. Methods Phys. Res. A, and T. LeCompte and H.T. Diehl, 
%{\sl The CDF and D0 Upgrades for Run II},  Ann. Rev. Nucl. Part. Sci. 
%{\bf 50}, 71 (2000).

\bibitem{foot1}
The pseudorapidity is defined as a function
of the polar angle $\theta$ as $\eta \equiv - \ln(\tan{\frac{\theta}{2}})$. 


%\bibitem{emu-lumi}
%Luminosity ID group: { http://www-d0.fnal.gov/phys\_id/luminosity/data\_access/ }


\bibitem{ttbar-prd}
D0 Collaboration, V.M. Abazov {\it et al.}, %V. Abazov {\it et al.}, 
submitted to Phys. Rev. D,
%Fermilab-Pub-07/128-E, 
arXiv:0705.2788.




%\bibitem{emu-T42-1}
%U. Bassler and G. Bernardi, 
%Toward a coherent treatment of calorimeter energies, D\O -Note 4124


\bibitem{emu-pythia} T.~Sjostrand {\sl et al.},
% P.~Eden, C.~Friberg, L.~Lonnblad,
%G.~Miu, S.~Mrenna and E.~Norrbin, 
%{\sl High-energy-physics event
%generation with  {\sc pythia},} 
Comput.\ Phys.\ Commun.\ {\bf 135}, 238 (2001). 

\bibitem{emu-CTEQ} H.~L.~Lai {\it et al.}, 
%{\sl Improved Parton Distributions 
%from Global Analysis of Recent Deep Inelastic Scattering and Inclusive Jet Data,} 
Phys. Rev. D55 (1997) 1280.

\bibitem{emu-COMPHEP} A. Pukhov {\sl et al.}, 
    %%CompHEP - a package for evaluation of Feynman diagrams and 
    %%integration over multi-particle phase space, 
[arXiv:hep-ph/9908288] (1999).

\bibitem{emu-ALPGEN}
M. Mangano {\it et al.},[arXiv:hep-ph/0206293] (2002).
%{\sl  ALPGEN, a generator for hard multiparton processes 
%in hadron collisions,} 



    %%\bibitem{emu-syst}
    %% S. Beauceron {\it et al.}, Systematic Studies Towards a new WH $\rar e \nu b \bar{b}$
    %%Cross Section Limit, D0 Note 4383 


    %%\bibitem{emu-bb}
    %%S. Beauceron and G. Bernardi, A Search  for $W b \bar{b}$ and $WH$ Production at $\sqrt{s}=1.96$~TeV / long note,
    %% D0 Note 4394

    %%\bibitem{emu-jlip1}
    %%D. Bloch {\it et al.}, Jet Lifetime $b$ Tagging, D0 Note 4069
    %%\bibitem{emu-jlip2}
    %%D. Bloch {\it et al.}, Performance of the JLIP $b$-tagger in p14, 
    %%D0 Note 4348 (see also D0 Notes 4158,4159)


\bibitem{emu-mcfm} 
J. Campbell and K. Ellis, {\sc mcfm} , {\sl Montecarlo for FeMtobarn processes},  http://mcfm.fnal.gov/

%\bibitem{emu-GEANT} Y.~Fisyak and J.~Womersley, D\O -Note 3191.


\bibitem{geant}
R.~Brun and F.~Carminati, CERN Program Library Long Writeup
      W5013, 1993 (unpublished).


\bibitem{blazey}  G. Blazey {\sl et al.}, in {\sl Proceedings of the  
workshop ``QCD and Weak Boson Physics in Run II''} edited by 
U. Baur, R.K. Ellis, and D. Zeppenfeld, Batavia
(2000), p. 47.


\bibitem{emu-JLIP05}
S. Greder, 
%{\sl   $b$ jet identification at D\O }, in preparation. 
{\sl $b$ quark tagging and cross-section measurement in quark pair production
at D\O },  FERMILAB-THESIS-2004-28 (2004).

%\bibitem{emu-foot3}{The variable $H_T$ is defined
%as the scalar sum of the transverse momentum of the jets.}

% \bibitem{emu-d0-notezbb}
%  D0 Collaboration, V.M. Abazov {\it et al.},
% {   \sl Evidence for  $Z \rar b\bar{b}$ production the D0 detector,}
%  D0 Conference-Note xxxx-CONF,\\ 
% { http://www-d0.fnal.gov/Run2Physics/WWW/results/prelim/HIGGS/H12/H12.pdf}

\bibitem{emu-junkLim} T. Junk, 
Nucl. Instrum. and Methods A {\bf 434}, 435 (1999).  


%%%%%%%%%%%%%%%%%%%%% WADE \begin{thebibliography}{99}
% \bibitem{emu-leplim}
% R.~Barate {\it et al.}  [LEP Working Group for Higgs boson searches],
%  %``Search For The SM Higgs Boson At Lep,''
%  Phys.\ Lett.\ B {\bf 565}, 61 (2003),
%  [arXiv:hep-ex/0306033]


\bibitem{emu-wadeLim} W. Fisher, FERMILAB-TM-2386-E (2007).

\bibitem{emu-smooth} K. S. Cranmer, Comput. Phys. Commun. {\bf 136}, 198
(2001) [arXiv:hep-ph/0011057].



% \bibitem{emu-fv} The procedure of replacing very low statistics MC sample
% shapes with a higher statistics version was performed for the
% following background/analysis combinations: $W/Z+2$jets
% ($WH \rightarrow e,\mu \nu b\bar{b}$); $Z\rightarrow e^+e^-, \tau^+
% \tau^-$ ($H\rightarrow e^+\nu e^-\nu$)

\bibitem{emu-interp} A. Read, 
Nucl. Instrum. and Methods  A {\bf 425}, 357 (1999).



%\bibitem{emu-pflh} W. Fisher, D\O -Note 5309.

\bibitem{emu-shwg}M. Carena {\it et al.}, [arXiv:hep-ph/0010338] (2000).

\bibitem{sm-lep}
ALEPH, DELPHI, L3 and OPAL Collaborations, The LEP Working Group for Higgs Boson Searches,
Phys. Lett. B
{\bf 565}, 61 (2003).


    %% 
    %% 
    %% \bibitem{emu-Bert} 
    %% I. Bertram {\sl et al.}, 
    %% Fermilab-TM-2104 (2000)
    %% 
    %% \bibitem{emu-csip} 
    %% R. Demina, A. Khanov and F. Rizatdinova, $b$ tagging with Counting Signed Impact Parameter Method, D\O -Note 4049
    %% \bibitem{emu-svt}
    %% A. Schwartzman and M. Narain, $b$ quark jet identification via secondary vertex reconstruction, D\O -Note 4080
    %% 
    %% \bibitem{emu-avto}
    %% A. Haas, A. Kharchilava and G. Watts,
    %% A D0 Search for Neutral Higgs Bosons at High $\tan \beta$ in Multi--jet Events, D\O - Note 4366
    %% 
    %% 
    %% 
    %% \bibitem{emu-T42-2}
    %%    \rm J.R. Vlimant {\it et al.},
    %%    \sl Technical description of the T42 algorithm for the calorimeter noise
    %%     suppression
    %%    \rm D\O -Note 4146, 2003.
    %% 
    %% 
    %% 
    %% 
    %% 
    %% %\bibitem[Bea04]{d0-note4394}
    %% \bibitem{emu-d0-note4633}
    %%      S. Beauceron and G. Bernardi,
    %%    \sl Search for $Wb\bar{b}$ and $WH$ Production in $p\bar{p}$ Collisions at $\sqrt{s}=1.96\,\mbox{TeV}$,
    %%  %\rm D0 Note 4394
    %%  \rm D0 Note 4633
    %% 
    %% \bibitem{emu-topanalyze}
    %%      {\tt http://www-d0.fnal.gov/Run2Physics/top/index.html}
    %% 
    %% \bibitem{emu-d0-CS-Pass2}
    %%      {\tt http://www-d0.fnal.gov/Run2Physics/cs/skimming/pass2.html}
    %% 
    %% \bibitem{emu-d0-note4350}
    %%      C. Cl\'ement {\it et al.},
    %%    \sl MuonID Certification for p14
    %%  \rm D0 Note 4350
    %% 
    %% 
    %% \bibitem{emu-d0-note4900}
    %%    C. Clement {\it et al.},
    %%   \sl Measurement of the $t\bar{t}$ production cross section at $\sqrt{s}=1.96$~TeV using lifetime tagging
    %%   \rm D0 Note 4900
    %% 
    %% 
    %% \bibitem{emu-d0-skim}
    %%      {\tt http://www-d0.fnal.gov/Run2Physics/cs/skimming/skimming.html}
    %% 
    %% \bibitem {d0-subskim}
    %%      {\tt http://www-d0.fnal.gov/Run2Physics/top/d0\_private/wg/commonskims/ data\_rootuples\_Ipanema.html}
    %% 
    %% 
    %% \bibitem{emu-jetmet}
    %%      {\tt http://www-d0.fnal.gov/~d0upgrad/d0\_private/software/ jetid/jetid.html}
    %% 
    %% \bibitem{emu-dqcalo}
    %%         S. Shary and L. Duflot,
    %%         {\sl DQ-calo, a package for the calorimeter data quality}
    %% 
    %% 
    %% \bibitem{emu-pythia62}
    %%    T. Sj\"ostrand {\sl et al.},
    %%    \sl {\tt pythia} 6.2: PHYSICS AND MANUAL,
    %%    \rm e-Print Archive: hep-ph/0108264, LU-TP-01-21, Lund 2001.
    %% 
    %% \bibitem{emu-CTEQ} 
    %%    H.~L.~Lai {\sl et al.},
    %%   \sl Improved Parton Distributions from Global Analysis of Recent Deep Inelastic Scattering and Inclusive Jet Data, 
    %%   \rm Phys. Rev. {\bf D}55 (1997) 1280
    %% 
    %% \bibitem{emu-mlm}
    %%    M. Mangano {\it et al.},
    %%    \sl {\tt ALPGEN}, a generator for hard multiparton processes in hadron collisions,
    %%    \rm {\tt http://mcfm.fnal.gov}.
    %% 
    %% \bibitem{emu-comphep}
    %%    A. Pukhov {\it et al.},
    %%    \sl {\tt CompHEP}, a package for evaluation of Feynman diagrams and integration over multi-particle phase space,
    %%    \rm INP-MSU98-41/542, Moscow 1999.
    %% 
    %% \bibitem{emu-mcfm}
    %%    J. Campbell and K. Ellis,
    %%    \sl {\tt mcfm}, Monte-Carlo for FeMtobarn processes,
    %%    \rm {\tt http://mcfm.fnal.gov}.
    %% 
    %% 
    %% \bibitem{emu-dogstar}
    %%    \rm Y. Fisyak and J. Womersley, 
    %%    \rm D\O -Note 3191
    %% 
    %% \bibitem{emu-geant}
    %%   \rm M. Goossens {\sl et al.},
    %%   \sl {\tt GEANT}, Detector Description and Simulation Tool,
    %%   \rm GEANT user's guide CERN, Geneva 1994.
    %% 
    %% \bibitem {d0-luminosity}
    %%      {\tt http://www-d0.fnal.gov/phys\_id/luminosity/data\_access/lm\_access}
    %% 
    %% 
    %% \bibitem{emu-luminosity_ID}
    %%       {\sl Luminosity ID group},
    %%      {\tt http://www-d0.fnal.gov/phys\_id/luminosity/ data\_access/doc}
    %% 
    %% \bibitem{emu-golling}
    %%      T. Golling,
    %%    \sl Measurement of the $t\bar{t}$ Production Cross-Section at
    %%        $\sqrt{s}=1.96\,$TeV in the Muon+Jets Final State using a
    %%        Topological Method
    %%    \rm D0 Conference note version 1.0, 2004.
    %% 
    %% \bibitem{emu-gollub}
    %%      N. Gollub {\it et al.},
    %%    \sl Measurement of the $t\bar{t}$ Production Cross-Section at
    %%        $\sqrt{s}=1.96\,$TeV in the Muon+Jets Final State using a
    %%        Topological Method on 363/pb of PASS2 data
    %%    \rm D\O -Note 4954, 2005.
    %% 
    %% \bibitem{emu-T42-1}
    %%    \rm U. Bassler and G. Bernardi,
    %%    \sl Towards a coherent treatment of calorimeter energies, 
    %%    \rm D\O -Note 4124, 2002.
    %% 
    %% \bibitem{emu-LP03}
    %%    \rm S. Beauceron, G. Bernardi and S. Trincaz-Duvoid, 
    %%    \sl Towards a Measurement of the $W+2$jets and $Wb\bar{b}$ Production Cross Section 
    %%        at $sqrt{s}=1.96\,$TeV   
    %%    \rm D\O -Note 4224, 2003.
    %% 
    %% \bibitem{emu-T42-3}
    %%    \rm G. Bernardi, E. Busato and J.R. Vlimant,
    %%    \sl Improvements from the T42 Algorithm on Calorimeter Objects Reconstruction, 
    %%    \rm D\O -Note 4335, 2004.
    %% 
    %% 
    %% 
    %%    
    %% %\bibitem{emu-james}
    %% %   \rm J. Heinmiller, 
    %% %  \sl Jet reconstruction efficiency as a function of $p_{\perp}$, for the Higgs group,
    %% %   \rm May 27${^th}$ 2004.
    %% 
    %% \bibitem{emu-JES53}
    %%    \rm Jet Energy Scale Group,
    %% {\tt http://www-d0.fnal.gov/phys\_id/jes/d0\_private/ certified/v5.3/jetcorr\_v5.3.html}.
    %% 
    %% 
    %% \bibitem{emu-moriond-04}
    %%    \rm S. Beauceron and G. Bernardi, 
    %%    \rm D0 Conference-Note 4399-CONF,\\ {\tt http://www-d0.fnal.gov/Run2Physics/WWW/results/HIGGS/H05/H05.pdf}
    %% 
    %% \bibitem{emu-hyunwoo-new}
    %%      H. Kim and J. Yu,
    %%    \sl Update on Search for $Wb\bar{b}$ and $WH$ Production in $p\bar{p}$ Collisions at $\sqrt{s}=1.96\,\mbox{TeV}$ with Full Data Set of Pass2, for the 2006 Winter conference.
    %% 
    %% 
    %% 
    %% 
    %% 
    %% %\bibitem{emu-d0-note4900}
    %% %     C. Cl\'ement {\it et al.},
    %% %   \sl Measurement of the ttbar cross section using Lifetime Tagging.
    %% % \rm D0 Note 4900
    %% 
    %% 
    %% %\bibitem[CDF99]{cdfsearch}
    %% %  \rm F. Abe {\it et al.},
    %% %  \sl Search for new particles decaying to $b\bar{b}$ in $p\bar{p}$ collisions wt $\sqrt{s}=1.8\,$TeV,
    %% %  Phys. Rev. Lett. 82 (1999), 2038.
    %% 
    %% 
    %% \bibitem{emu-d0-note4264}
    %%      D. Bauer, J. Huang and A. Zieminski,
    %%    \sl Studies of Upsilon state production with the D0 detector at FNAL
    %%  \rm D0 Note 4264
    %% 
    %% 
    %% \bibitem{emu-JetID}
    %%    \rm B. Andrieu,
    %%    {\tt http://www-d0.fnal.gov/phys\_id/jets/jetid.html}.
    %% 
    %% 
    %% \bibitem{emu-private}
    %%    \rm S. Choi, K. Hanagaki, 
    %%    \sl , private comunication.
    %% 
    %% \bibitem{emu-Bert} 
    %%    \rm I. Bertram {\sl et al.}, 
    %%        Fermilab-TM-2104 (2000).
    %% 
    %% 
    %% %\bibitem{emu-klute}
    %% %   \rm M. Klute,
    %% %   \sl Measurement of the ttbar cross section at $sqrt{s}=1.96\,$TeV 
    %% %       in muon-plus-jets channel
    %% %   \rm D\O -Note 4185, 2003.
    %% 
    %% %\bibitem[Bus03]{busato}
    %% %   \rm E. Busato,
    %% %   {\tt http://www-d0.fnal.gov/d0upgrad/d0\_private/software/jetid/meetings2003/Jul08/busato.ppt}.
    %% 
    %% 
    %% \bibitem{emu-d0-note4988}
    %%    A. Jenkins {\it et al.},
    %%   \sl Evidence for  $b\bar{b}$ decays at D0
    %%   \rm D0 Note 4900
    %% 
    %% \bibitem{emu-avto}
    %%    \rm A. Haas, A. Kharchilava, G. Watts,
    %%    \sl A D0 Search for Neutral Higgs Bosons at High $\tan\beta$ in Multi-jet Events,
    %%    \rm D\O -Note 4366, 2004.
    %% 




\end{thebibliography}
\end{document}

\begin{tabular}{lcccccccccccc}
\hline
{Higgs mass [GeV]}                  & 100 & 110    & 120    & 130    & 140    & 150    & 160    & 170    & 180    & 190    & 200  \\
\hline\hline
$WH  \rightarrow  WWW$ observed $\sigma \times B$/pb &   --   & 71.41 &  11.27 &   4.41 &   1.57 &   0.52 &   0.09 &  0.018 &  0.010 &  0.006 &  0.004 \\
$WH  \rightarrow  WWW$ expected $\sigma \times B$/pb &   --   & 65.35 &  10.78 &   3.53 &   1.30 &   0.41 &   0.07 &  0.013 &  0.007 &  0.004 &  0.003 \\
\hline
$H  \rightarrow  WW$ observed $\sigma \times B$/pb &  10.79 &   5.61 &   6.07 &   5.94 &   4.24 &   4.06 &   3.69 &   4.38 &   4.07 &   3.50 &   3.25 \\
$H  \rightarrow  WW$ expected $\sigma \times B$/pb &   8.94 &   6.31 &   7.74 &   6.18 &   5.25 &   4.52 &   3.58 &   3.68 &   3.40 &   3.28 &   3.98 \\
\hline
\hline
$WH  \rightarrow  WWW$ observed ratio to SM&   --   & 455.9 &  110.7 &   74.7 &   53.7 &   46.0 &   46.1 &   53.5 &   62.1 &   75.5 &  89.6 \\
$WH  \rightarrow  WWW$ expected ratio to SM&   --   & 417.2 &  105.9 &   59.8 &   44.7 &   36.4 &   34.4 &   38.1 &   44.6 &   50.1 &  60.8 \\
\hline
$H  \rightarrow  WW$ observed ratio to SM&  636.4 &   98.9 &   46.1 &   26.4 &   14.0 &   11.8 &    9.9 &   13.3 &   15.4 &   19.1 &   22.2 \\
$H  \rightarrow  WW$ expected ratio to SM&  527.5 &  111.2 &   58.8 &   27.4 &   17.3 &   13.1 &    9.6 &   11.2 &   12.8 &   17.9 &   27.2 \\
\hline
\hline
%$CL_s$ observed $\sigma \times B$/pb &   2.05 &   2.20 &   3.16 &   4.50 &   4.15 &   3.30 &   3.47 &   4.07 &   3.80 &   3.21 &   3.09 \\
%$CL_s$$ expected $\sigma \times B$/pb &   2.71 &   2.82 &   3.54 &   4.68 &   5.14 &   4.41 &   3.20 &   3.33 &   3.15 &   2.91 &   3.67 \\
%\hline
%$CL_s$ observed ratio to SM&    5.4 &    7.2 &   10.6 &   14.0 &   11.8 &   9.1  &    9.2 &   12.3 &   14.3 &   17.5 &   21.0 \\
%$CL_s$ expected ratio to SM&    7.1 &    9.1 &   11.9 &   14.6 &   14.6 &  12.1  &    8.5 &   10.1 &   11.9 &   15.8 &   25.0 \\
\hline
D0 observed ratio to SM&    5.5 &    7.1 &   10.5 &   14.2 &   12.8 &   10.2 &   10.2 &   13.7 &   16.1 &   20.7 &   23.7 \\
D0  expected ratio to SM&    8.7 &   10.8 &   14.3 &   15.7 &   13.8 &   11.7 &    9.0 &   10.5 &   12.1 &   16.0 &   23.5 \\
\hline
\end{tabular}